\documentclass[a4paper,11pt]{article}
\pdfoutput=1 

\usepackage{jheppub} 
\usepackage[T1]{fontenc} 
\usepackage[utf8]{inputenc}
\usepackage{physics}
\usepackage{amsfonts,shuffle}
\usepackage{yfonts} 
\newcommand{\uproman}[1]{\uppercase\expandafter{\romannumeral#1}}
\def\bdelta{\boldsymbol{\delta}}
\usepackage{nicefrac}
\usepackage{mathrsfs}
\usepackage{mathtools}
\usepackage{empheq}
\usepackage{shuffle}
\usepackage{standalone}
\usepackage{subcaption} 
\usepackage{cleveref}
\usepackage{tikz} 
\usetikzlibrary{fadings}
\usetikzlibrary{patterns}
\usetikzlibrary{shadows.blur}
\usetikzlibrary{shapes}
\usepackage{mathrsfs}
\usepackage{subcaption}
\usepackage{adjustbox}
\usepackage{bigints} 

\newdimen\satlevel
\newdimen\satdiameter
\newcommand{\satisfaction}[2][]{%
    \satdiameter=2.5ex\relax
    \ifcase#2\relax
        \satlevel=0pt\relax
    \or
        \satlevel=0.125\satdiameter
    \or
        \satlevel=0.25\satdiameter
    \or
        \satlevel=0.31\satdiameter
    \or
        \satlevel=0.5\satdiameter
    \fi
    \tikz[baseline=-0.3\satdiameter]{%
        \draw[#1,  line width=1.5] (0,0) circle (0.5\satdiameter);
        \fill[#1] (0,0) circle (\satlevel);
    }%
    }
\usepackage{xcolor} 
\definecolor{CelestialRed}{HTML}{D0021B}
\definecolor{CelestialGreen}{HTML}{00B531}

\usepackage{bm}
 
\preprint{\vbox{\hbox{\hphantom{XXX}LMU-ASC 11/24}}}
\title{\boldmath Celestial String Integrands \& their Expansions}

\author[]{Daniel Bockisch}

\affiliation[]{Arnold-Sommerfeld-Center for Theoretical Physics, \\ 
Ludwig-Maximilians-Universität, 80333 München, Germany}

\emailAdd{Daniel.Bockisch@physik.uni-muenchen.de}

\abstract{
We transform the one-loop four-point type \MakeUppercase{\romannumeral 1} open superstring gluon amplitude to correlation functions on the celestial sphere including both the (non-)orientable planar and non-planar sector. This requires a Mellin transform with respect to the energies of the scattered strings, as well as to integrate over the open-string worldsheet moduli space. After accomplishing the former we obtain celestial string integrands with remaining worldsheet integrals $\Psi\left(\beta\right)$, where $\beta$ is related to the conformal scaling dimensions of the conformal primary operators under consideration. Employing an alternative approach of performing an $\alpha'$-expansion of the open superstring amplitude first and Mellin transforming afterwards, we obtain a fully integrated expression, capturing the pole structure in the $\beta$-plane. The same analysis is performed at tree-level yielding similar results. We conclude by solving $\Psi\left(\beta\right)$ for specific values of $\beta$, consistently reproducing the results of the $\alpha'$-expansion ansatz. In all approaches we find that the dependence on $\alpha'$ reduces to that of a simple overall factor of $\left(\alpha'\right)^{\beta-3}$ at loop and $\left(\alpha'\right)^{\beta}$ at tree level, consistent with previous literature.
}

\begin{document} 
\maketitle
\flushbottom

\newpage

\section{Introduction}

The celestial holography program aims to establish a duality between quantum gravity in four-dimensional asymptotically flat bulk spacetimes and a two-dimensional (celestial) conformal field theory (CCFT) living on its asymptotic boundary at null infinity \cite{stromingerLecturesInfraredStructure2018}, the eponymous celestial sphere. Developing and working within the already existing preliminary framework of CCFT is an active area of research, see for instance chapter 3 of \cite{pasterskiCelestialHolography2021a} and references therein for an overview.

Scattering amplitudes in four-dimensional flat spacetime play a central role in constructing the CCFT from the bottom-up as they are the holographic dual to two-dimensional conformal correlators on the celestial sphere \cite{ pasterskiFlatSpaceAmplitudes2017, pasterskiConformalBasisFlat2017, pasterskiGluonAmplitudes2d2017, stiebergerSymmetriesCelestialAmplitudes2019}, more commonly referred to as celestial amplitudes. 

For massless particles, the transition from flat space amplitudes $\mathcal{A}$ to celestial amplitudes $\Tilde{\mathcal{A}}$ is explicitly furnished by a Mellin transform
\begin{align}
	\mathcal{\tilde A}(\Delta_i) \equiv \prod_{i=1}^{n} \int_{0}^{\infty} \dd{\omega_{i}} \omega_{i}^{\Delta_i-1} \mathcal{A}\left(\omega_{i} \right). 
 \label{eq:BasicCelestialAmplitude}
\end{align}
It  effectively exchanges each of the energies $\omega_{i}$ of the scattered particles with 
 the corresponding dual variable, the conformal scaling dimension $\Delta_{i} $ associated to a conformal primary operator $\mathcal{O}_{\Delta_{i}}$ in the CCFT.

The prospect of understanding flat space holography from the bottom-up by performing well-posed integrals makes for a great motivation to compute celestial amplitudes. It is therefore not surprising that a respectable amount of literature has been published in recent years successfully pursuing precisely this approach. Examples at tree-level include, with no claim of completeness, gluon amplitudes in pure YM-theory at $3$ and $4$ points \cite{pasterskiGluonAmplitudes2d2017}, at $n$-points \cite{schreiberTreelevelGluonAmplitudes2018}, gluon-graviton and pure graviton amplitudes at $4$ points in Einstein-Yang-Mills and Einstein-gravity theory respectively \cite{stiebergerStringsCelestialSphere2018}, for $\mathcal{N}=4$ SYM-theory \cite{jiangCelestialSuperamplitudeMathcal2021, brandhuberCelestialSuperamplitudes2021} and last but not least four point type \MakeUppercase{\romannumeral 1} open and heterotic superstring amplitudes \cite{stiebergerStringsCelestialSphere2018}. At one-loop level, celestial amplitudes have been investigated in field theory \cite{gonzalezLoopCorrectionsCelestial2020, albayrakLoopCelestialAmplitudes2020} and for open superstrings \cite{donnayCelestialOpenStrings2023a}.

In this paper we want to continue the analysis of celestial string amplitudes initiated at tree level in \cite{stiebergerStringsCelestialSphere2018} and recently continued at loop level in \cite{donnayCelestialOpenStrings2023a}. There are some appealing reasons to study string theory amplitudes in particular in this context. In general, string theory amplitudes possess salient features that set them apart from regular field theory amplitudes and make them particularly well suited and interesting to be studied in celestial holography. One of them is the well-known ultra soft high-energy behavior of string amplitudes \cite{grossStringTheoryPlanck1988}, rendering them UV finite. This is a particularly beneficial feature when computing celestial amplitudes which, by their very definition via the Mellin transform (\ref{eq:BasicCelestialAmplitude}), are sensitive to the entire energy scale from IR to UV. Indeed, these anti-Wilsonian characteristics of celestial amplitudes and their implications have been highlighted on general grounds in \cite{arkani-hamedCelestialAmplitudesUV2021} and will naturally make an appearance in this work as well. 
\begin{figure}[t]
   \centering
  \includegraphics[width=0.55\textwidth]{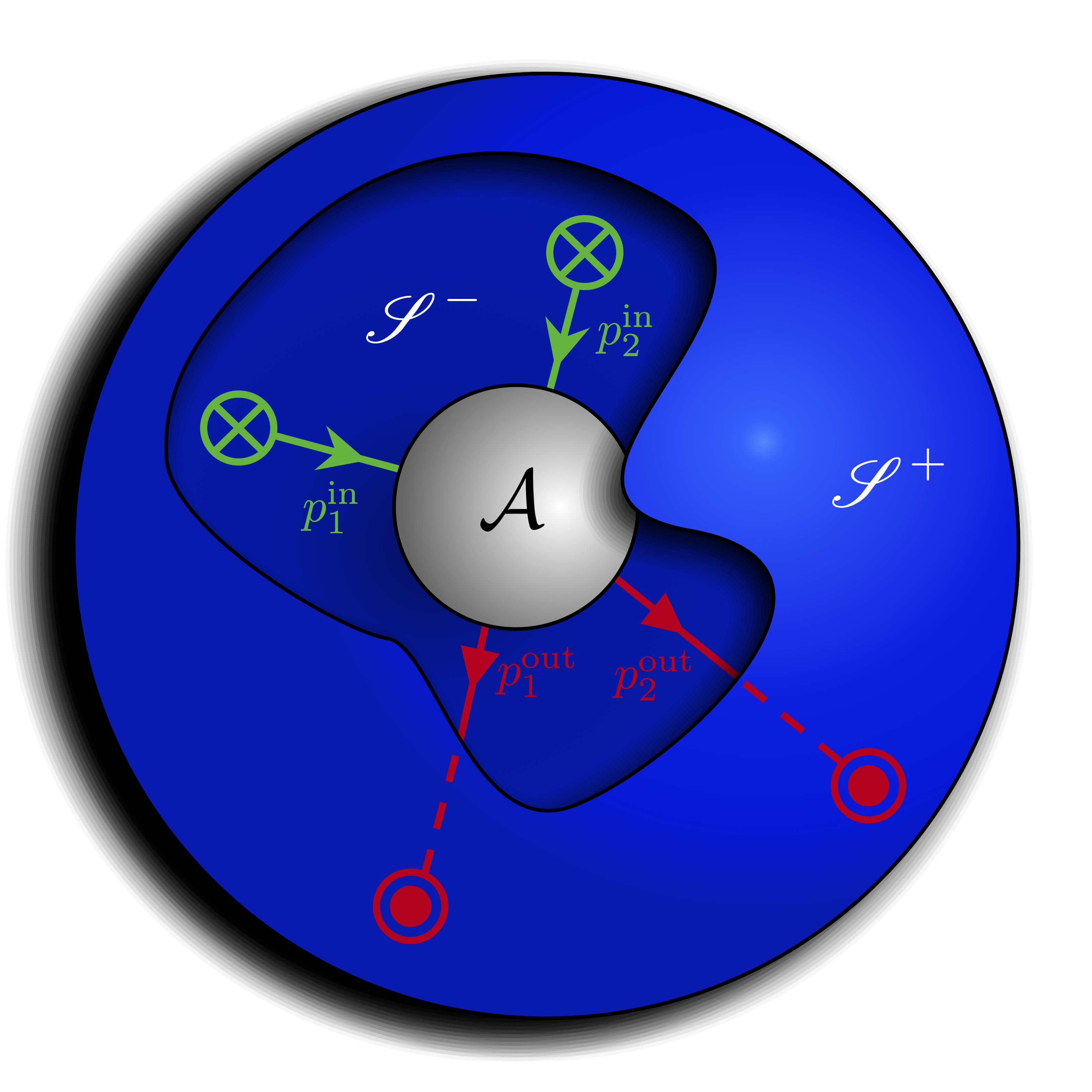}
   \caption{Conceptual visualization of the correspondence between amplitudes in asymptotically flat space and conformal correlators on the celestial sphere. Depicted here is a $2 \to 2$ scattering process. The ingoing states enter the spacetime from past null infinity $\mathscr{Script\ font}^{\! \! -}$ marked at $\color{CelestialGreen} \LARGE \pmb{\otimes} $, the scattering occurs in the bulk and the resulting outgoing states eventually exit the bulk at future null infinity $\mathscr{Script\ font}^{\! \! +}$ marked with $\satisfaction[CelestialRed]{3}$ .}   
   \label{fig:CelestialSphereBulkBoundaryPicture}
\end{figure}
Another more conceptual motivation for studying string amplitudes in the celestial context is the remarkable similarity between the geometric pictures of string and celestial scattering amplitudes. In celestial holography the in and out states involved in a scattering process in Minkowski space are exiting and eventually entering past and future null infinity respectively. This is where the two-dimensional celestial sphere resides and consequently the point of entrance and exit are seen as the CCFT operator insertion points cf.  
figure \ref{fig:CelestialSphereBulkBoundaryPicture}, in what is essentially the celestial incarnation of a state-operator correspondence \cite{crawleyStateoperatorCorrespondenceCelestial2021a}.

To anyone knowledgeable in string scattering amplitudes this picture seems suspiciously familiar, as tree level string scattering amplitudes can be defined as CFT operator insertions in the two-dimensional worldsheet which has been conformally mapped to a sphere. One might be tempted to write this off as a superficial similarity. However, Stieberger and Taylor were able to show that at four-points, the string world sheet indeed can be mapped to the celestial sphere in the celestial equivalent of the ultra-high energy limit  \cite{stiebergerStringsCelestialSphere2018}. This analysis was extended very recently to five-points \cite{castiblancoCelestialStringsField2024}. Indeed, there has been some excellent works successfully applying (ambitwistor) string theory methods in celestial holography \cite{buSupersymmetricCelestialOPEs2021, jiangCelestialOPEsW_2021, adamoCelestialOperatorProducts2021, adamoCelestialW_Infty2022, casaliCelestialDoubleCopy2020, adamoCelestialAmplitudesConformal2019}, including the first top-down toy model for celestial holography \cite{costelloTopdownHolographyAsymptotically2023, costelloBurnsSpaceHolography2023a}.

The paper is organized as follows. With the goal of being self-contained, we review and collect all the basic definitions and objects from celestial holography and string scattering needed throughout the rest of the paper in section \ref{sec:Preliminaries} and \ref{sec:GeneralStringAmplitudes} respectively. In section \ref{sec:CelestialTrees} we give a very brief summary of celestial string amplitudes at tree level, which represents the first successful fusion of the previous two topics. In section \ref{sec:CelestialLoopIntegrands} we begin extending this analysis by computing celestial string loop integrands of the four-point one-loop type \MakeUppercase{\romannumeral 1} open superstring amplitude for both the planar orientated annulus, the not orientable M{\"o}bius strip as well as the non-planar contribution by performing a Mellin transform with respect to the energies of the scattered strings. Performing the remaining integrals over the string worldsheet coordinates and the usual modular parameter $ q = e^{2\pi i\tau} $ after the Mellin transform turns out to be highly non-trivial. Section \ref{sec:ExpansionCelestialLoopAmplitudes} aims to circumvents this difficulty by using well-known string amplitude expansions in the inverse string tension $\alpha'$. While this ansatz does not produce a closed-form expression, it nevertheless captures some important features of celestial string amplitudes, namely their pole structure in the complex plane of the conformal scaling dimension, and allows for the integration over the string moduli, ultimately yielding to a fully integrated expression. As a proof of concept we first perform the expansion at tree level where a closed-form expression of the corresponding celestial string amplitude is known and compare the results. With this intuition at hand we perform the same analysis at loop-level where the closed-form expression is not known. We conclude by providing an explicit connection between these results and the ones from the previous section.

\section{Preliminaries \& Conventions}
\label{sec:Preliminaries}

This section serves mainly to introduce and fix our notation, which will adhere to the ones found in \cite{stiebergerStringsCelestialSphere2018}. Throughout this paper, we will focus exclusively on four point scattering amplitudes of gluons and their stringy incarnations for simplicity. Consequently, expressions presented in the following are tailored to this purpose, though generalizations to higher points are of course principally possible. 

As seen in equation (\ref{eq:BasicCelestialAmplitude}), computing celestial amplitudes requires defining regular momentum space amplitudes. To this end, we parameterize the four-momentum of a massless particle as
\begin{align}
	p^{\mu} = \omega q^{\mu} = \frac{1}{2}\left(1+z\bar{z},z+\bar{z},-i\left(z+\bar{z}\right),1-z\bar{z}\right),
 \label{eq:FourMomenta}
\end{align}
where $\omega$ is the energy and $z,\bar{z}$ are complex coordinates parametrizing the celestial sphere. Correspondingly, the spinor helicity brackets read
\begin{align}
    \langle ij \rangle = \sqrt{\omega_{i}\omega_{j}} \, z_{ij} \, , \qquad 
     [ij] = -\sqrt{\omega_{i}\omega_{j}} \, \bar{z}_{ij} \,  ,
\end{align}
where $z_{ij} \equiv z_{i}-z_{j}$ as usual. The spinor-helicity-formalism allows us to express amplitudes in an incredibly compact manner as exemplified by the famous Parke-Taylor formula (PT) for tree-level MHV gluon amplitudes. We will only consider polarisation configurations $A\left(1^{-},2^{-},3^{+},4^{+}\right)$ throughout the rest of this paper. At four-points, the PT formula then simply reads
\begin{align}
    A_{\text{YM}}^{\text{tree}}(1^-,2^-,3^+,4^+) = \frac{\langle 12\rangle^3}{\langle 23\rangle \langle 34\rangle \langle 41\rangle}.
    \label{eq:ParkeTaylor}
\end{align}
Here we stripped off the group theoretic color factors and the momentum conserving delta distribution. We will denote amplitudes including said distribution as
\begin{align}
    \mathcal{A}(1,2,3,4) = A(1,2,3,4) \, \delta^{(4)}\left(p\right).
\end{align}
The Mandelstam variables expressed in the spinor-helicity formalism read 
\begin{align}
    s_{ij} \equiv 2p_{i}p_{j} = \langle ij\rangle[ji]  =  \omega_{i}\omega_{j}z_{ij}\bar{z}_{ij}.     \label{eq:MandelstamVariable}
\end{align} 
In the case of four-point scattering there are generally three independent Mandelstam variables because
\begin{align}
    s_{12}=s_{34}, \qquad s_{13}=s_{24}, \qquad s_{14}=s_{23}.
    \label{eq:MandelstamEquivalents}
\end{align}
However, since we deal exclusively with massless scattering 
\begin{align}
    s_{12} + s_{13} + s_{23} = \sum_{i} m_{i}^{2} = 0, 
    \label{eq:MandelstamMasslessCondition}
\end{align}
the system is further reduced to two independent Mandelstam variables. We will only use $s_{12} \equiv s $ and $s_{23} \equiv -u $ for the remainder of this work. 
Since the four point amplitudes we are interested in will turn into four point correlators on the celestial sphere, it is advisable as well as customary to write it as a product of a prefactor that satisfies conformal covariance and objects that are invariant under conformal transformations. This object is the conformal cross ratio 
\begin{align}
    r = \frac{z_{12}z_{34}}{z_{23}z_{41}},
\end{align}
which can be related to a ratio of the aforementioned Mandelstam variables and consequently to the the scattering angle $\theta$ in the center of mass frame 
\begin{align}
    \frac{s_{23}}{s_{12}} = - \frac{u}{s} = \frac{1}{r} = \sin^2{\left( \frac{\theta}{2}\right)}.
    \label{eq:ConformalCrossRatio}
\end{align}
It also makes an appearance when expressing the momentum conserving delta function in terms our four momenta (\ref{eq:FourMomenta}), yielding 
\begin{align}
    \delta^{(4)}\left(p \right) \equiv  \delta^{(4)}\left(\sum_{j=1}^{4}\varepsilon_{j}\omega_{j}q_{j} \right) 
    = \frac{4}{\omega_{4}(z_{14}\bar{z}_{14})(z_{23}\bar{z}_{23})}\delta(r-\bar{r}) \prod_{i=1}^{3} \delta\left(\omega_i - \chi_{i} \, \omega_{4}\right),
    \label{eq:MomentumDeltaCelestialCoordinates}
\end{align}
thereby fixing all Mellin integrals except for the one over $\omega_{4}$, where we defined
\begin{align}
    \chi_{1} = \frac{z_{24}\bar{z}_{34}}{z_{12}\bar{z}_{13}}, \qquad \chi_{2} = -\frac{z_{14}\bar{z}_{34}}{z_{12}\bar{z}_{23}}, \qquad \chi_{3} = -\frac{z_{24}\bar{z}_{14}}{z_{23}\bar{z}_{13}}.
\end{align}
Note that $\delta(r-\bar{r})$ in (\ref{eq:MomentumDeltaCelestialCoordinates}) constraints the conformal cross ratio $r$ to be real. This constraint together with (\ref{eq:ConformalCrossRatio}) immediately yields $r>1$ which will be encoded in our equations as a heavyside function $\theta(r-1)$. The aforementioned kinematic prefactor takes the form
\begin{align}    
K\left(h_i,\bar{h}_i\right) = \prod_{i<j}^{4} z_{ij}^{h/3-h_i-h_j}\bar{z}_{ij}^{\bar{h}/3-\bar{h}_i-\bar{h}_j},
\end{align}
where $h= \sum_{i=1}^{4} h_{i}$ and $\bar{h}= \sum_{i=1}^{4} \bar{h}_{i}$ are the sums over the conformal weight $h_{i}$ of the CCFT operator associated to the i-th scattered particle. These can be used to define the conformal scaling dimension $\Delta_{i}$ and the 2d spin of the particle $J_{i}$ in the usual CFT manner
\begin{align}
    \Delta_{i} = h_{i} + \bar{h}_{i}, \qquad J_{i} = h_{i} - \bar{h}_{i}.
\end{align}    
The spins are predetermined to match exactly the four dimensional helicities of the original amplitude. Additionally the conformal scaling dimension need to lie on the principal continous series $\Delta_{i} = 1 + i\lambda_{i}$ where $\lambda_{i} \in \mathbb{R}$ \cite{pasterskiConformalBasisFlat2017}. While these conditions do not completely fix the conformal weights, a possible choice which we will adhere to is 
\begin{equation}
\begin{aligned}
    &h_1 = \frac{i}{2}\lambda_1, \quad &&h_2 = \frac{i}{2}\lambda_2, \quad &&h_3 = 1 + \frac{i}{2}\lambda_3, \quad &&h_4 = 1 + \frac{i}{2}\lambda_4,   \\ 
    &\bar{h}_1 = 1 + \frac{i}{2}\lambda_1, \quad &&\bar{h}_2 = 1 + \frac{i}{2}\lambda_2 \quad &&\bar{h}_3 = \frac{i}{2}\lambda_3 \quad &&\bar{h}_4 =  \frac{i}{2}\lambda_4. 
\end{aligned}
\end{equation}
A quantity of interest related to these conformal weights is (the sum of) their imaginary part. To this end we introduce  
\begin{align}
    \beta \equiv -\frac{i}{2} \sum_{i=1}^{4}\lambda_{i},
    \label{eq:beta}
\end{align}
which also allows us to write the sum of all conformal scaling dimensions as
\begin{align}
    \Delta \equiv \sum_{i=1}^{4}\Delta_{i} = 4 + i \sum_{i=1}^{4}\lambda_{i} = 4 - 2\beta. 
    \label{eq:SumOfDelta}
\end{align}
We have now assembled all the ingredients necessary to define a four point celestial amplitude as 
\begin{align}
	\mathcal{\tilde{A}} \left(\{ \Delta_{i} ,  \, z_{i} , \, \bar{z}_{i}    \} \right) \equiv  \prod_{i=1}^{4} \int_{0}^{\infty} \dd{\omega_{i}} \omega_{i}^{\Delta_i-1} \mathcal{A}\left(\{ \omega_{i} ,  \, z_{i} , \, \bar{z}_{i}    \} \right). 
 \label{eq:CelestialAmplitude}
\end{align}

\section{String Amplitudes}
\label{sec:GeneralStringAmplitudes}

For the purposes of this paper we are primarily interested in the expansion of open superstring amplitudes in terms of the inverse string tension $\alpha'$. Note that since $\alpha'$ is a dimensional quantity, it needs to be accompanied by appropriate powers of Mandelstam variables $s_{ij}$ to render the overall expansion parameter dimensionless. It is for this very reason that some authors choose to absorb the $\alpha '$ directly into the Mandelstam variables $s_{ij}$. However, to keep the notation consistent with section \ref{sec:Preliminaries} and emphasise a feature of celestial string amplitudes later on, we choose to keep those explicitly separated.

\begin{figure}[t]
    \centering
    $\mathcal{A}_{\text{Open}} = $
    $\raisebox{-1.ex}{$\bigint_{\mathcal{M}_{0,4}}$}  $ 
    \begin{adjustbox}{scale=0.25}
        \tikzset{every picture/.style={line width=0.75pt}} 

\begin{tikzpicture}[x=0.75pt,y=0.75pt,yscale=-1,xscale=1,baseline={([yshift=-3ex]current bounding box.center)}]

\draw  [fill={rgb, 255:red, 173; green, 173; blue, 173 }  ,fill opacity=1 ][line width=0.75]  (183.17,162.08) .. controls (183.17,87.11) and (243.94,26.33) .. (318.92,26.33) .. controls (393.89,26.33) and (454.67,87.11) .. (454.67,162.08) .. controls (454.67,237.06) and (393.89,297.83) .. (318.92,297.83) .. controls (243.94,297.83) and (183.17,237.06) .. (183.17,162.08) -- cycle ;
\draw  [fill={rgb, 255:red, 0; green, 0; blue, 0 }  ,fill opacity=1 ] (425.88,66.05) .. controls (425.88,74.27) and (419.22,80.92) .. (411,80.92) .. controls (402.79,80.92) and (396.13,74.27) .. (396.13,66.05) .. controls (396.13,57.84) and (402.79,51.18) .. (411,51.18) .. controls (419.22,51.18) and (425.88,57.84) .. (425.88,66.05) -- cycle ;
\draw  [fill={rgb, 255:red, 0; green, 0; blue, 0 }  ,fill opacity=1 ] (220.68,90.85) .. controls (220.68,99.07) and (214.02,105.72) .. (205.8,105.72) .. controls (197.59,105.72) and (190.93,99.07) .. (190.93,90.85) .. controls (190.93,82.64) and (197.59,75.98) .. (205.8,75.98) .. controls (214.02,75.98) and (220.68,82.64) .. (220.68,90.85) -- cycle ;
\draw  [fill={rgb, 255:red, 0; green, 0; blue, 0 }  ,fill opacity=1 ] (289.87,289.25) .. controls (289.87,297.47) and (283.22,304.12) .. (275,304.12) .. controls (266.79,304.12) and (260.13,297.47) .. (260.13,289.25) .. controls (260.13,281.04) and (266.79,274.38) .. (275,274.38) .. controls (283.22,274.38) and (289.87,281.04) .. (289.87,289.25) -- cycle ;
\draw  [fill={rgb, 255:red, 0; green, 0; blue, 0 }  ,fill opacity=1 ] (457.48,215.25) .. controls (457.48,223.47) and (450.82,230.12) .. (442.6,230.12) .. controls (434.39,230.12) and (427.73,223.47) .. (427.73,215.25) .. controls (427.73,207.04) and (434.39,200.38) .. (442.6,200.38) .. controls (450.82,200.38) and (457.48,207.04) .. (457.48,215.25) -- cycle ;

\end{tikzpicture}
    \end{adjustbox} 
    \text{\Large{+}} \hspace{-1em} $ \raisebox{-1.ex}{ $\bigint_{\mathcal{M}_{1,4}} $} \Bigg( $
    \begin{adjustbox}{scale=0.25}
        \tikzset{every picture/.style={line width=0.75pt}} 

\begin{tikzpicture}[x=0.75pt,y=0.75pt,yscale=-1,xscale=1,baseline={([yshift=-3ex]current bounding box.center)}]

\draw  [fill={rgb, 255:red, 173; green, 173; blue, 173 }  ,fill opacity=1 ][line width=0.75]  (318.92,26.33) .. controls (393.89,26.33) and (454.67,87.11) .. (454.67,162.08) .. controls (454.67,237.06) and (393.89,297.83) .. (318.92,297.83) .. controls (243.94,297.83) and (183.17,237.06) .. (183.17,162.08) .. controls (183.17,87.11) and (243.94,26.33) .. (318.92,26.33) -- cycle (235.63,162.08) .. controls (235.63,208.08) and (272.92,245.38) .. (318.92,245.38) .. controls (364.92,245.38) and (402.21,208.08) .. (402.21,162.08) .. controls (402.21,116.08) and (364.92,78.79) .. (318.92,78.79) .. controls (272.92,78.79) and (235.63,116.08) .. (235.63,162.08) -- cycle ;
\draw  [fill={rgb, 255:red, 0; green, 0; blue, 0 }  ,fill opacity=1 ] (395.08,44.85) .. controls (395.08,53.07) and (388.42,59.72) .. (380.2,59.72) .. controls (371.99,59.72) and (365.33,53.07) .. (365.33,44.85) .. controls (365.33,36.64) and (371.99,29.98) .. (380.2,29.98) .. controls (388.42,29.98) and (395.08,36.64) .. (395.08,44.85) -- cycle ;
\draw  [fill={rgb, 255:red, 0; green, 0; blue, 0 }  ,fill opacity=1 ] (210.68,105.65) .. controls (210.68,113.87) and (204.02,120.52) .. (195.8,120.52) .. controls (187.59,120.52) and (180.93,113.87) .. (180.93,105.65) .. controls (180.93,97.44) and (187.59,90.78) .. (195.8,90.78) .. controls (204.02,90.78) and (210.68,97.44) .. (210.68,105.65) -- cycle ;
\draw  [fill={rgb, 255:red, 0; green, 0; blue, 0 }  ,fill opacity=1 ] (231.48,254.85) .. controls (231.48,263.07) and (224.82,269.72) .. (216.6,269.72) .. controls (208.39,269.72) and (201.73,263.07) .. (201.73,254.85) .. controls (201.73,246.64) and (208.39,239.98) .. (216.6,239.98) .. controls (224.82,239.98) and (231.48,246.64) .. (231.48,254.85) -- cycle ;
\draw  [fill={rgb, 255:red, 0; green, 0; blue, 0 }  ,fill opacity=1 ] (438.67,248.05) .. controls (438.67,256.27) and (432.02,262.92) .. (423.8,262.92) .. controls (415.59,262.92) and (408.93,256.27) .. (408.93,248.05) .. controls (408.93,239.84) and (415.59,233.18) .. (423.8,233.18) .. controls (432.02,233.18) and (438.67,239.84) .. (438.67,248.05) -- cycle ;

\end{tikzpicture}
    \end{adjustbox}
    \text{\Large{+}} \hspace{-0.66em}
    \begin{adjustbox}{scale=0.25}
 \raisebox{-3.3cm}{\includegraphics[width=0.55\textwidth]{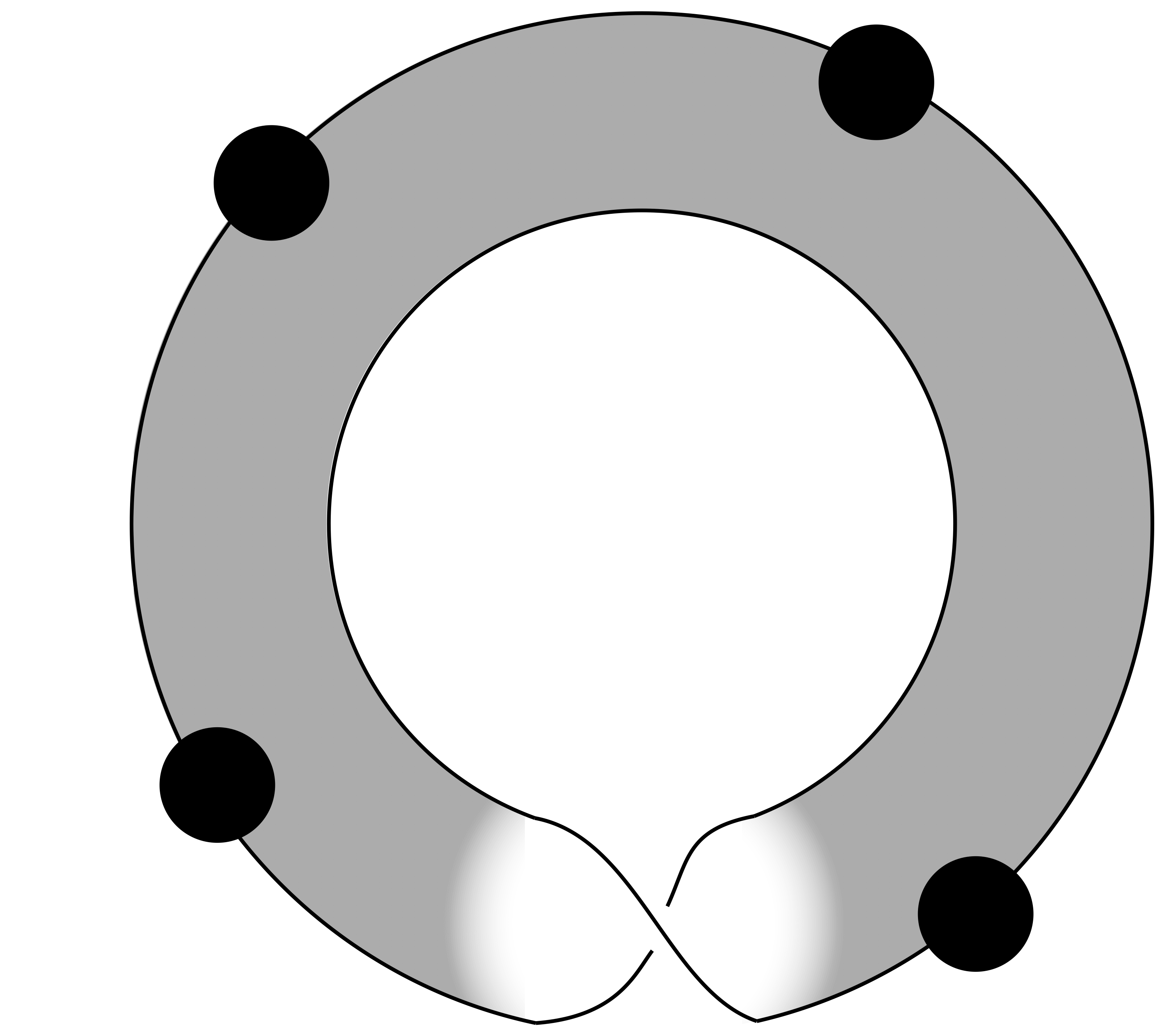}}
    \end{adjustbox}
        \text{\Large{+}}
    \begin{adjustbox}{scale=0.25}
        \tikzset{every picture/.style={line width=0.75pt}} 

\begin{tikzpicture}[x=0.75pt,y=0.75pt,yscale=-1,xscale=1,baseline={([yshift=-2.45ex]current bounding box.center)}]

\draw  [fill={rgb, 255:red, 173; green, 173; blue, 173 }  ,fill opacity=1 ][line width=0.75]  (318.92,26.33) .. controls (393.89,26.33) and (454.67,87.11) .. (454.67,162.08) .. controls (454.67,237.06) and (393.89,297.83) .. (318.92,297.83) .. controls (243.94,297.83) and (183.17,237.06) .. (183.17,162.08) .. controls (183.17,87.11) and (243.94,26.33) .. (318.92,26.33) -- cycle (235.63,162.08) .. controls (235.63,208.08) and (272.92,245.38) .. (318.92,245.38) .. controls (364.92,245.38) and (402.21,208.08) .. (402.21,162.08) .. controls (402.21,116.08) and (364.92,78.79) .. (318.92,78.79) .. controls (272.92,78.79) and (235.63,116.08) .. (235.63,162.08) -- cycle ;
\draw  [fill={rgb, 255:red, 0; green, 0; blue, 0 }  ,fill opacity=1 ] (406.68,124.55) .. controls (406.68,132.77) and (400.02,139.42) .. (391.8,139.42) .. controls (383.59,139.42) and (376.93,132.77) .. (376.93,124.55) .. controls (376.93,116.34) and (383.59,109.68) .. (391.8,109.68) .. controls (400.02,109.68) and (406.68,116.34) .. (406.68,124.55) -- cycle ;
\draw  [fill={rgb, 255:red, 0; green, 0; blue, 0 }  ,fill opacity=1 ] (259.38,52.85) .. controls (259.38,61.07) and (252.72,67.72) .. (244.5,67.72) .. controls (236.29,67.72) and (229.63,61.07) .. (229.63,52.85) .. controls (229.63,44.64) and (236.29,37.98) .. (244.5,37.98) .. controls (252.72,37.98) and (259.38,44.64) .. (259.38,52.85) -- cycle ;
\draw  [fill={rgb, 255:red, 0; green, 0; blue, 0 }  ,fill opacity=1 ] (269.17,211.45) .. controls (269.17,219.67) and (262.52,226.32) .. (254.3,226.32) .. controls (246.09,226.32) and (239.43,219.67) .. (239.43,211.45) .. controls (239.43,203.24) and (246.09,196.58) .. (254.3,196.58) .. controls (262.52,196.58) and (269.17,203.24) .. (269.17,211.45) -- cycle ;
\draw  [fill={rgb, 255:red, 0; green, 0; blue, 0 }  ,fill opacity=1 ] (333.79,297.83) .. controls (333.79,306.05) and (327.13,312.7) .. (318.92,312.7) .. controls (310.7,312.7) and (304.05,306.05) .. (304.05,297.83) .. controls (304.05,289.62) and (310.7,282.96) .. (318.92,282.96) .. controls (327.13,282.96) and (333.79,289.62) .. (333.79,297.83) -- cycle ;

\end{tikzpicture}
    \end{adjustbox}
    $\Bigg) $ \text{\Large{+}} $\cdots$ 
    \caption[]{Expansion of the four point open string amplitude in terms of moduli space integrals over punctured Riemann surfaces up to and including genus one. The punctures are represented by solid black dots \tikz[x=0.75pt,y=0.75pt,yscale=-0.22,xscale=0.22]
{\draw  [fill={rgb, 255:red, 0; green, 0; blue, 0 }  ,fill opacity=1 ] (331.58,161.65) .. controls (331.58,169.87) and (324.92,176.52) .. (316.7,176.52) .. controls (308.49,176.52) and (301.83,169.87) .. (301.83,161.65) .. controls (301.83,153.44) and (308.49,146.78) .. (316.7,146.78) .. controls (324.92,146.78) and (331.58,153.44) .. (331.58,161.65) -- cycle ;}. Accompanying string coupling constants $g_{s}$, gauge group generators and permutations over the ordering of the punctures have been suppressed for simplicity.}
    \label{fig:OpenStringScatteringToOneLoopOrderConceptual}
\end{figure}

 The rich mathematical underpinning of string scattering amplitudes allows the full amplitude to be expanded in terms of integrals over the moduli spaces Riemann surfaces $\mathcal{M}_{g,n}$ of genus $g$ with $n$ punctures positioned on their respective boundaries. The genus $g$ classifies the loop-level of the contributions whereas the number of punctures $n$ correspond to the $n$-point scattering amplitudes. We will restrain ourselves to tree $(g=0)$ and one loop $(g=1)$ level. At tree level the only contribution to the open string amplitude is the disk. At one loop level there are three contributions. Two of those are orientable and captured by the annulus with the punctures either all inserted on one or inserted on both boundaries. These are referred to as planar and non-planar respectively. The remaining contribution is represented by the famously non orientable M{\"o}bius strip with all punctures inserted on its only boundary. A visual summary of this picture can be seen in Figure \ref{fig:OpenStringScatteringToOneLoopOrderConceptual}. In the following subsections we will present the amplitudes resulting from this approach.   
\subsection{String Tree Amplitudes}
\label{sec:StringTreeAmplitudes}

It has been known for a long time that type \MakeUppercase{\romannumeral 1} superstring tree amplitudes at four points can be written as a product of the pure YM gluon tree level amplitude and a string formfactor \cite{Green:1981xx} 
\begin{align}
    \mathcal{A}_{\text{\MakeUppercase{\romannumeral 1}}}^{\text{tree}}(1,2,3,4) = g_{10}^{2} \,  \mathcal{A}_{\text{YM}}^{\text{tree}}(1,2,3,4) \,  F_{I}(\alpha' s,\alpha' u),
    \label{eq:StringTreeAmplitude}
\end{align}
which can be expressed in terms of the Beta and by extension Gamma functions
\begin{align}
    F_{I}(\alpha' s,\alpha' u) = -\alpha' s_{12} \,  \mathcal{B}\left(-\alpha' s_{12}, 1 + \alpha' s_{23} \right) = \frac{\Gamma(1-\alpha' s_{12})\Gamma(1+\alpha's_{23})}{\Gamma(1-\alpha's_{12}+\alpha's_{23})}. 
\end{align}
The formfactor allows for a Laurent expansion in $\alpha'$, ultimately leading to 
\begin{align}
F_{I}(\alpha' s,\alpha' u)  
    &= \exp{
    \sum_{k=2}^{\infty} (-\alpha')^k \frac{\zeta_{k}}{k}\left[
    (-s_{12})^k + s_{23}^k - (-s_{12} + s_{23})^k
    \right]
    } \nonumber \\ 
&= 1 + \left(\alpha'\right)^2\zeta_{2}s_{12}s_{23} + \left(\alpha'\right)^3\zeta_{3}s_{12}s_{23}(s_{12}-s_{23}) \nonumber \\ 
&\phantom{=} + \left(\alpha'\right)^4\zeta_{4}s_{12}s_{23}\left(s_{12}^2-\frac{1}{4}s_{12}s_{23}+s_{23}^2
\right) + \mathcal{O}\left(\alpha'^5\right),
\label{eq:ExpansionVeneziano} 
\end{align}
where the coefficients $\zeta_{n}$ appearing in this expansion are Riemann zeta values\footnote{The condition $n\geq 2$ in (\ref{eq:RiemannZetaValue}) is needed for convergence as the case $n=1$ yields the divergent harmonic series.} 
\begin{align}
     \zeta_{n} \equiv \sum_{k=1}^{\infty} \frac{1}{k^{n}} , \quad n\in \mathbb{N}_{\geq 2}.
    \label{eq:RiemannZetaValue}
\end{align}
The first term in (\ref{eq:ExpansionVeneziano}) yields the pure YM term while all terms proportional to $\alpha'$ encode the string contributions to the amplitude. Note that at tree level, there is no linear term in $\alpha'$. This will become relevant when computing the celestial amplitude. 
The appearance of zeta values persists for gluon tree level amplitudes in superstring theory to arbitrary higher orders and for $N$ points \cite{mafraCompleteNPointSuperstring2011a, mafraCompleteNPointSuperstring2011}, albeit in a generalized form, known as multiple Zeta Values (MZVs). They are defined via the convergent series
\begin{align}
    \zeta\left(a_{1},a_{2},\ldots, a_{k}\right) =  \sum_{n_1 > \ldots > n_k >0} \, \prod_{j=1}^{k} \,  n_j^{-a_{j}}, \quad a_{j} \in \mathbb{N}_{\geq 1}, \quad a_{1} \geq 2.
    \label{eq:GeneralMZVOfDepthK}
\end{align}
MZVs are classified by the total number of arguments $k$ and the sum $\sum_{j}^{k} a_{j}$ called \emph{depth} and \emph{weight} respectively. In light of this, Riemann zeta values $\zeta_{n}$ can be interpreted as MZVs of depth $1$ and weight $n$. Note that in the $\alpha'$-expansion (\ref{eq:ExpansionVeneziano}) the weight of the MZV always matches the total power of the accompanying Mandelstam variables \cite{broedelPolylogarithmsMultipleZeta2013}.
The reason for the appearance of MZV coefficients in the $\alpha'$-expansion lies in the aforementioned underlying integration over the moduli space of punctured Riemann surfaces of genus zero, i.e. the disk. It turns out that MZVs are the periods of iterated integrals over these spaces  \cite{brownMultipleZetaValues2009}. 
\subsection{String Loop Amplitudes}
\label{sec:StringLoopAmplitudes}

\begin{figure}[t]
    \centering
    \vspace{2em}
    \begin{subfigure}[b]{0.46\textwidth}
        \resizebox{\textwidth}{!}{
        \includegraphics[width=\textwidth]{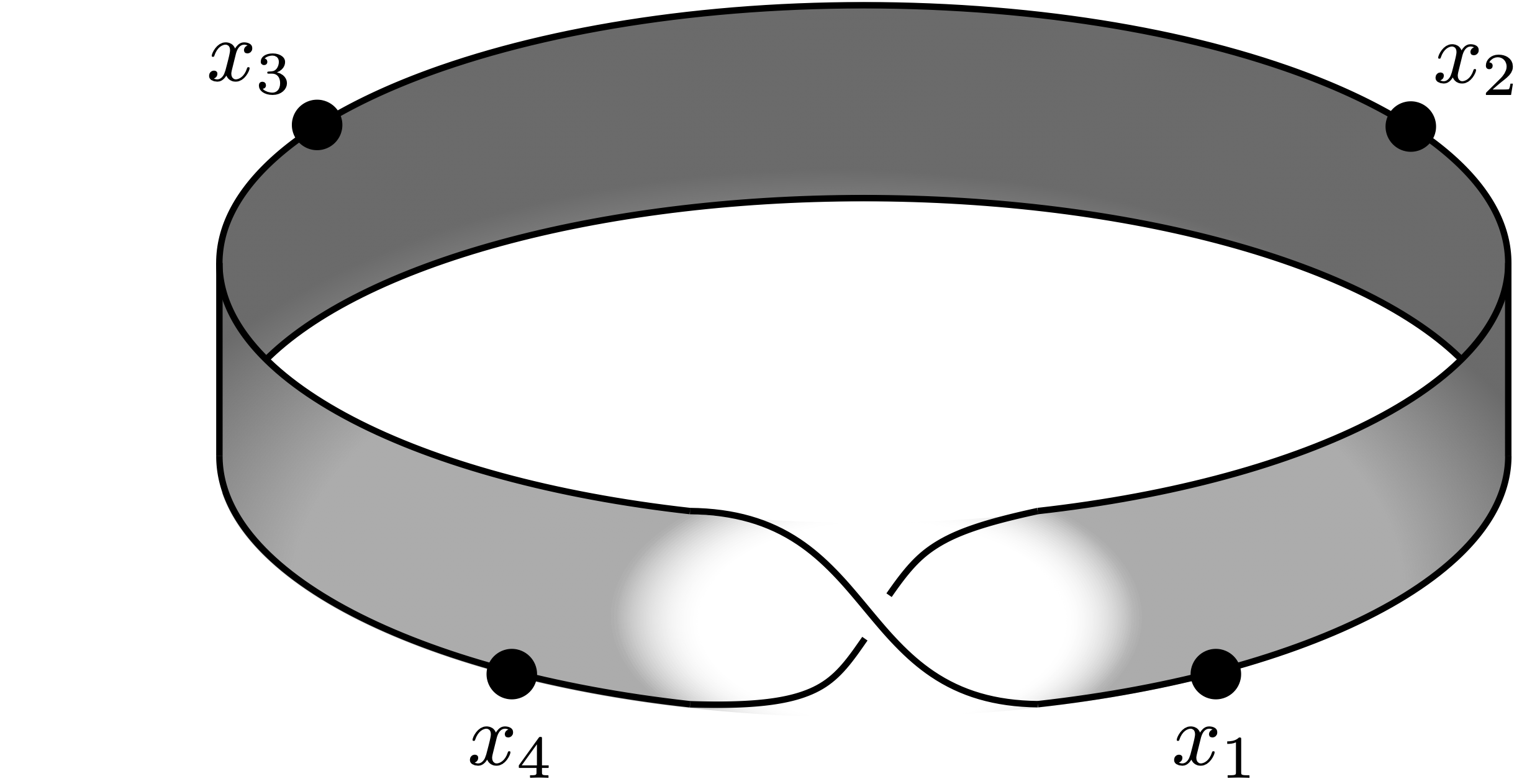}                  
        }
        \caption{M{\"o}bius strip}
        \label{fig:3DMoebiusStrip}
    \end{subfigure}
    \hspace{-2em}
    \begin{subfigure}[b]{0.57\textwidth}
        \resizebox{\textwidth}{!}{
        \includegraphics[width=\textwidth]{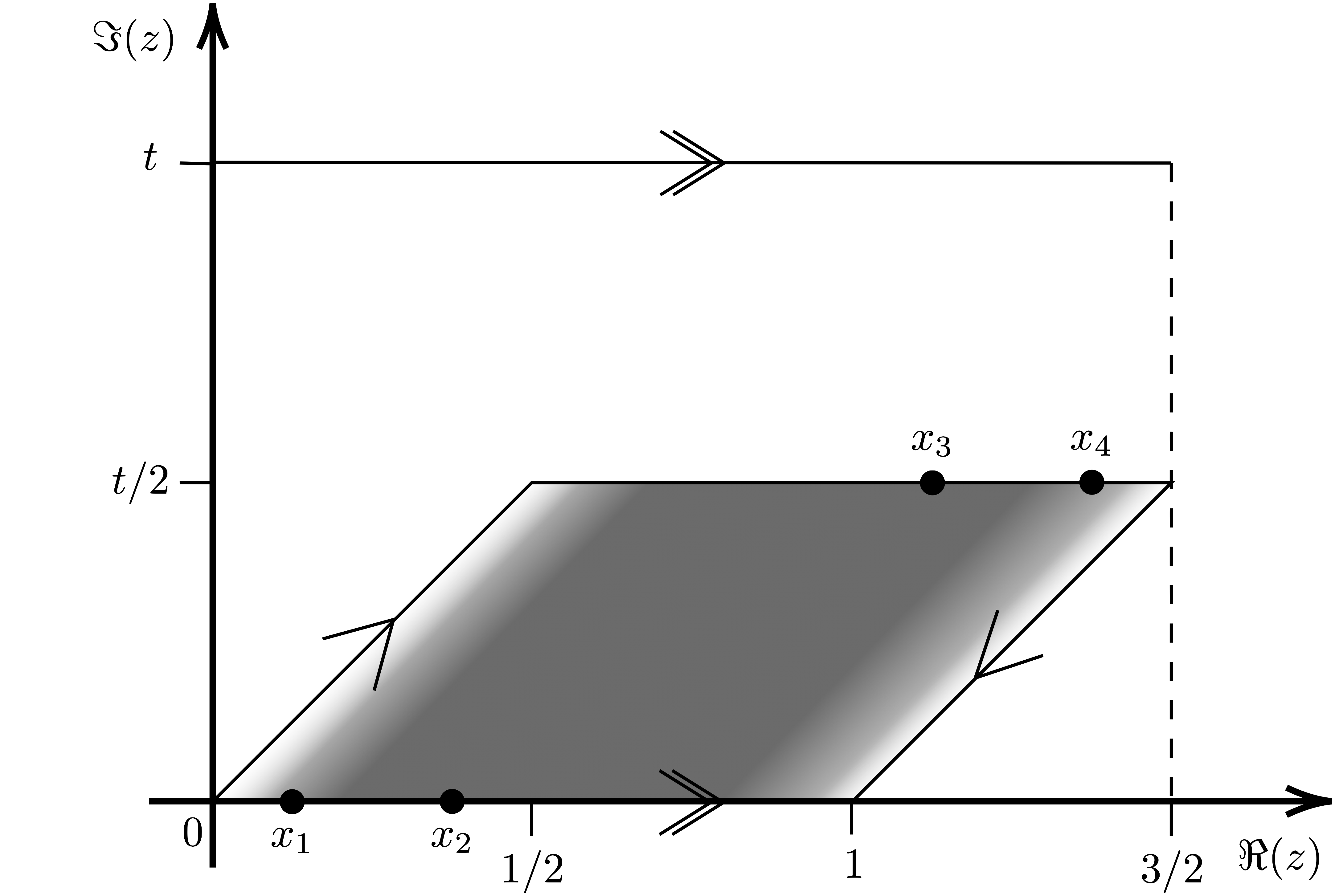}
        }
        \caption{M{\"o}bius strip worldsheet.}
        \label{fig:WorldsheetMoebiusStrip}
    \end{subfigure}
    \caption[]{
    Non-orientable M{\"o}bius strip contribution to open string scattering possessing only a \emph{single} boundary consisting of the horizontal lines at $\imaginary{(z)}=0$ and $\imaginary{(z)}=t/2$. 
    }
    \label{fig:MoebiusStrip}
\end{figure}

With the conceptual Figure \ref{fig:OpenStringScatteringToOneLoopOrderConceptual} in mind, it becomes clear that the main challenge in transitioning from tree to one loop level is computing the periods over the cylinder\footnote{The cylinder is topologically equivalent to the annulus so we will use these words interchangeably.} and the M{\"o}bius strip as opposed to the disk. The corresponding computations are significantly more involved than their tree level counterpart. Nevertheless, recent years saw the publication of an excellent body of work \cite{broedelEllipticMultipleZeta2015, broedelTwistedEllipticMultiple2018b} analysing and successfully evaluating these amplitudes at the level of the $\alpha'$-expansion. A similar pattern as in tree level case emerges at one loop, the crucial difference being that the MZVs are replaced with their genus \emph{one} equivalent, the \emph{elliptical} multiple zeta values (eMZVs), see \cite{matthesOverviewEllipticMultiple2020} for some mathematical background.\footnote{It should be mentioned that the evaluation of the non-planar sector leads to the appearance of a slightly more sophisticated object called the \emph{twisted} elliptical multiple zeta value (teMZV). However, it has been shown \cite{broedelTwistedEllipticMultiple2018b} that in the $\alpha'$-expansion teMZVs can be reduced to a combination of eMZV and zeta values yet again.} In the following we will give a brief review how these objects materialise in the $\alpha'$-expansion of one loop open superstring amplitudes and ultimately write down the relevant expressions we aim to celestialize.
 \begin{figure}[t]
    \centering
    \begin{subfigure}[b]{0.49\textwidth}
        \resizebox{\textwidth}{!}{                  
  
\tikzset {_t021mxt2v/.code = {\pgfsetadditionalshadetransform{ \pgftransformshift{\pgfpoint{0 bp } { 0 bp }  }  \pgftransformrotate{-225 }  \pgftransformscale{2 }  }}}
\pgfdeclarehorizontalshading{_beyxeu6m6}{150bp}{rgb(0bp)=(0.8,0.8,0.8);
rgb(37.5bp)=(0.8,0.8,0.8);
rgb(47.76785714285714bp)=(0.68,0.68,0.68);
rgb(62.5bp)=(0.42,0.42,0.42);
rgb(100bp)=(0.42,0.42,0.42)}
\tikzset{every picture/.style={line width=0.75pt}} 

\begin{tikzpicture}[x=0.75pt,y=0.75pt,yscale=-1,xscale=1]
\path  [shading=_beyxeu6m6,_t021mxt2v][blur shadow={shadow xshift=0pt,shadow yshift=0pt, shadow blur radius=2.25pt, shadow blur steps=4 ,shadow opacity=100}] (236.32,119) -- (380.68,119) .. controls (396.04,119) and (408.5,141.39) .. (408.5,169) .. controls (408.5,196.61) and (396.04,219) .. (380.68,219) -- (236.32,219) .. controls (220.96,219) and (208.5,196.61) .. (208.5,169) .. controls (208.5,141.39) and (220.96,119) .. (236.32,119) .. controls (251.69,119) and (264.15,141.39) .. (264.15,169) .. controls (264.15,196.61) and (251.69,219) .. (236.32,219) ; 
 \draw   (236.32,119) -- (380.68,119) .. controls (396.04,119) and (408.5,141.39) .. (408.5,169) .. controls (408.5,196.61) and (396.04,219) .. (380.68,219) -- (236.32,219) .. controls (220.96,219) and (208.5,196.61) .. (208.5,169) .. controls (208.5,141.39) and (220.96,119) .. (236.32,119) .. controls (251.69,119) and (264.15,141.39) .. (264.15,169) .. controls (264.15,196.61) and (251.69,219) .. (236.32,219) ; 

\draw  [fill={rgb, 255:red, 0; green, 0; blue, 0 }  ,fill opacity=1 ] (213.89,143.66) .. controls (211.38,143.65) and (209.36,141.61) .. (209.37,139.1) .. controls (209.38,136.59) and (211.42,134.57) .. (213.93,134.58) .. controls (216.44,134.59) and (218.46,136.63) .. (218.45,139.14) .. controls (218.44,141.65) and (216.4,143.67) .. (213.89,143.66) -- cycle ;
\draw  [fill={rgb, 255:red, 0; green, 0; blue, 0 }  ,fill opacity=1 ] (258.6,143.84) .. controls (256.09,143.83) and (254.07,141.79) .. (254.08,139.28) .. controls (254.09,136.77) and (256.13,134.75) .. (258.64,134.76) .. controls (261.15,134.77) and (263.17,136.81) .. (263.16,139.32) .. controls (263.15,141.83) and (261.11,143.85) .. (258.6,143.84) -- cycle ;
\draw  [fill={rgb, 255:red, 0; green, 0; blue, 0 }  ,fill opacity=1 ] (258.36,203.97) .. controls (255.85,203.96) and (253.83,201.91) .. (253.84,199.41) .. controls (253.85,196.9) and (255.89,194.87) .. (258.4,194.88) .. controls (260.9,194.89) and (262.93,196.93) .. (262.92,199.44) .. controls (262.91,201.95) and (260.87,203.98) .. (258.36,203.97) -- cycle ;
\draw  [fill={rgb, 255:red, 0; green, 0; blue, 0 }  ,fill opacity=1 ] (213.65,203.79) .. controls (211.14,203.78) and (209.12,201.73) .. (209.13,199.23) .. controls (209.14,196.72) and (211.18,194.69) .. (213.69,194.7) .. controls (216.2,194.71) and (218.22,196.75) .. (218.21,199.26) .. controls (218.2,201.77) and (216.16,203.8) .. (213.65,203.79) -- cycle ;
\draw  [dash pattern={on 4.5pt off 4.5pt}] (352.85,169) .. controls (352.85,141.39) and (365.31,119) .. (380.68,119) .. controls (396.04,119) and (408.5,141.39) .. (408.5,169) .. controls (408.5,196.61) and (396.04,219) .. (380.68,219) .. controls (365.31,219) and (352.85,196.61) .. (352.85,169) -- cycle ;

\draw (207.37,139.1) node [anchor=east] [inner sep=0.75pt]    {$x_{1}$};
\draw (207.13,199.23) node [anchor=east] [inner sep=0.75pt]    {$x_{2}$};
\draw (265.16,139.32) node [anchor=west] [inner sep=0.75pt]    {$x_{4}$};
\draw (264.92,199.44) node [anchor=west] [inner sep=0.75pt]    {$x_{3}$};

\end{tikzpicture}
        }
        \caption{planar cylinder}
        \label{fig:3DPlanarCylinder}
    \end{subfigure}
    \hfill
    \begin{subfigure}[b]{0.49\textwidth}
        \resizebox{\textwidth}{!}{                
  
\tikzset {_bhis8aiqk/.code = {\pgfsetadditionalshadetransform{ \pgftransformshift{\pgfpoint{0 bp } { 0 bp }  }  \pgftransformrotate{-225 }  \pgftransformscale{2 }  }}}
\pgfdeclarehorizontalshading{_usdburhvr}{150bp}{rgb(0bp)=(0.8,0.8,0.8);
rgb(37.5bp)=(0.8,0.8,0.8);
rgb(47.76785714285714bp)=(0.68,0.68,0.68);
rgb(62.5bp)=(0.42,0.42,0.42);
rgb(100bp)=(0.42,0.42,0.42)}
\tikzset{every picture/.style={line width=0.75pt}} 

\begin{tikzpicture}[x=0.75pt,y=0.75pt,yscale=-1,xscale=1]

\path  [shading=_usdburhvr,_bhis8aiqk][blur shadow={shadow xshift=0pt,shadow yshift=0pt, shadow blur radius=2.25pt, shadow blur steps=4 ,shadow opacity=100}] (236.32,119) -- (380.68,119) .. controls (396.04,119) and (408.5,141.39) .. (408.5,169) .. controls (408.5,196.61) and (396.04,219) .. (380.68,219) -- (236.32,219) .. controls (220.96,219) and (208.5,196.61) .. (208.5,169) .. controls (208.5,141.39) and (220.96,119) .. (236.32,119) .. controls (251.69,119) and (264.15,141.39) .. (264.15,169) .. controls (264.15,196.61) and (251.69,219) .. (236.32,219) ; 
 \draw   (236.32,119) -- (380.68,119) .. controls (396.04,119) and (408.5,141.39) .. (408.5,169) .. controls (408.5,196.61) and (396.04,219) .. (380.68,219) -- (236.32,219) .. controls (220.96,219) and (208.5,196.61) .. (208.5,169) .. controls (208.5,141.39) and (220.96,119) .. (236.32,119) .. controls (251.69,119) and (264.15,141.39) .. (264.15,169) .. controls (264.15,196.61) and (251.69,219) .. (236.32,219) ; 

\draw  [fill={rgb, 255:red, 0; green, 0; blue, 0 }  ,fill opacity=1 ] (213.89,143.66) .. controls (211.38,143.65) and (209.36,141.61) .. (209.37,139.1) .. controls (209.38,136.59) and (211.42,134.57) .. (213.93,134.58) .. controls (216.44,134.59) and (218.46,136.63) .. (218.45,139.14) .. controls (218.44,141.65) and (216.4,143.67) .. (213.89,143.66) -- cycle ;
\draw  [fill={rgb, 255:red, 0; green, 0; blue, 0 }  ,fill opacity=1 ] (402.18,143.66) .. controls (399.67,143.65) and (397.65,141.61) .. (397.66,139.1) .. controls (397.67,136.59) and (399.71,134.57) .. (402.22,134.58) .. controls (404.73,134.59) and (406.75,136.63) .. (406.74,139.14) .. controls (406.73,141.65) and (404.69,143.67) .. (402.18,143.66) -- cycle ;
\draw  [fill={rgb, 255:red, 0; green, 0; blue, 0 }  ,fill opacity=1 ] (402.18,203.95) .. controls (399.67,203.94) and (397.65,201.9) .. (397.66,199.39) .. controls (397.67,196.88) and (399.71,194.85) .. (402.22,194.86) .. controls (404.73,194.87) and (406.75,196.92) .. (406.74,199.42) .. controls (406.73,201.93) and (404.69,203.96) .. (402.18,203.95) -- cycle ;
\draw  [fill={rgb, 255:red, 0; green, 0; blue, 0 }  ,fill opacity=1 ] (213.65,203.79) .. controls (211.14,203.78) and (209.12,201.73) .. (209.13,199.23) .. controls (209.14,196.72) and (211.18,194.69) .. (213.69,194.7) .. controls (216.2,194.71) and (218.22,196.75) .. (218.21,199.26) .. controls (218.2,201.77) and (216.16,203.8) .. (213.65,203.79) -- cycle ;
\draw  [dash pattern={on 4.5pt off 4.5pt}] (352.85,169) .. controls (352.85,141.39) and (365.31,119) .. (380.68,119) .. controls (396.04,119) and (408.5,141.39) .. (408.5,169) .. controls (408.5,196.61) and (396.04,219) .. (380.68,219) .. controls (365.31,219) and (352.85,196.61) .. (352.85,169) -- cycle ;

\draw (207.37,139.1) node [anchor=east] [inner sep=0.75pt]    {$x_{1}$};
\draw (207.13,199.23) node [anchor=east] [inner sep=0.75pt]    {$x_{2}$};
\draw (408.74,139.14) node [anchor=west] [inner sep=0.75pt]    {$x_{4}^{'}$};
\draw (408.74,199.42) node [anchor=west] [inner sep=0.75pt]    {$x_{3}^{'}$};

\end{tikzpicture}}
        \caption{non-planar cylinder}
        \label{fig:3DNonPlanarCylinder}
    \end{subfigure}
    \caption[]{
    String worldsheets with cylinder topology contributing to open string scattering.
    }
    \label{fig:Both3DCylinders}
\end{figure}
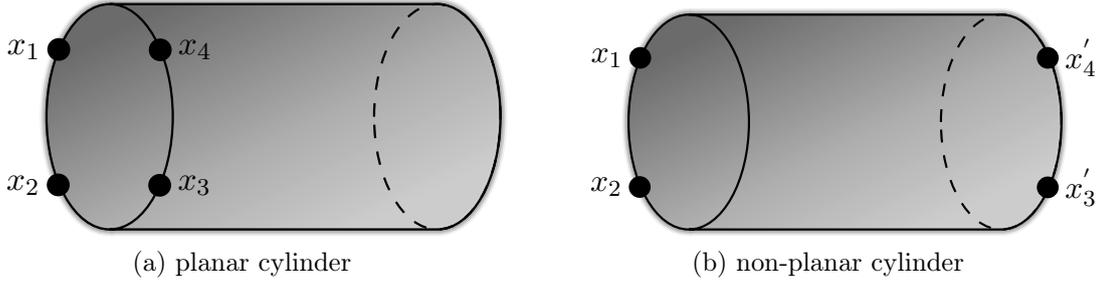
The first step involves mapping all relevant one loop worldsheets, i.e. the (non)-planar annulus and the M{\"o}bius strip presented in Figure \ref{fig:OpenStringScatteringToOneLoopOrderConceptual}, to the complex plane. This can be achieved for all cases by acting with different involutions on a covering torus $T_{2} = \mathbb{C}/\left(\mathbb{Z}+\mathbb{Z} \, \tau \right) $, i.e.\ a rectangle in the complex plane with periodic boundary conditions $z \equiv z + \tau$ and edge lengths $1$ and $\tau = it, \, t \in \mathbb{R}_{+} $ along the real and imaginary axis respectively. Here $\tau$ is the modular parameter of the torus $T_{2}$. In the case of both the planar and non-planar cylinder contribution depicted in Figure \ref{fig:Both3DCylinders}, the basic cylinder $C_{2}$ is obtained by folding in half the rectangle of the torus in the complex plane via the involution $I_{C}(z) = \bar{z}$, such that $C_{2}=T_{2}/I_{C}$, see Figure \ref{fig:BothCylinderWorldsheets}. Note that the horizontal lines $\imaginary{(z)}=0$ and $\imaginary{(z)}=t/2$ are fixed points with respect to the cylinder involution $I_{C}$ and define the two boundaries of the cylinder in which the string vertex operators are inserted. The M{\"o}bius strip $M_2$ can be obtained imposing an additional involution $I_{M}(z) = z + \tau/2 + 1/2 $ on the cylinder $M_{2} = C_{2}/I_{M}$. It identifies both cylinder boundaries and adds a twist, resulting in the desired non-orientable manifold with a single boundary, see Figure \ref{fig:MoebiusStrip}.\footnote{The exact form of these worldsheet pictures is highly dependent on convention. Here we mainly followed the conventions found in \cite{broedelTwistedEllipticMultiple2018b,gerkenModularGraphForms2020a}. For some alternative representations, see for instance \cite{nagaoInvolutiondependentConstantsCancellation1988,burgessOpenSuperstringsPolyakov1987,burgessOpenUnorientedStrings1987,angelantonjOpenStrings2002,vanhoveBuildingBlocksClosed2024}.}
The second step consists of setting up the integrals over the correlators on their respective worldsheets. To this end one employs the genus-one Green function 
\begin{align}
    P\left(z_i-z_j\right) = \frac{1}{2} \log\abs{\frac{\theta_1\left(z_i-z_j,\tau\right)}{\theta_1'\left(0,\tau\right)}}^2 - \frac{\pi}{\imaginary \left(\tau\right)} \left[\imaginary\left(z_i-z_j\right)\right]^2,
    \label{eq:GreenFunction}
\end{align}
\begin{figure}[t]
    \centering
    \hspace{-4em}
    \begin{subfigure}[b]{0.57\textwidth}
        \resizebox{\textwidth}{!}{
        \includegraphics[width=\textwidth]{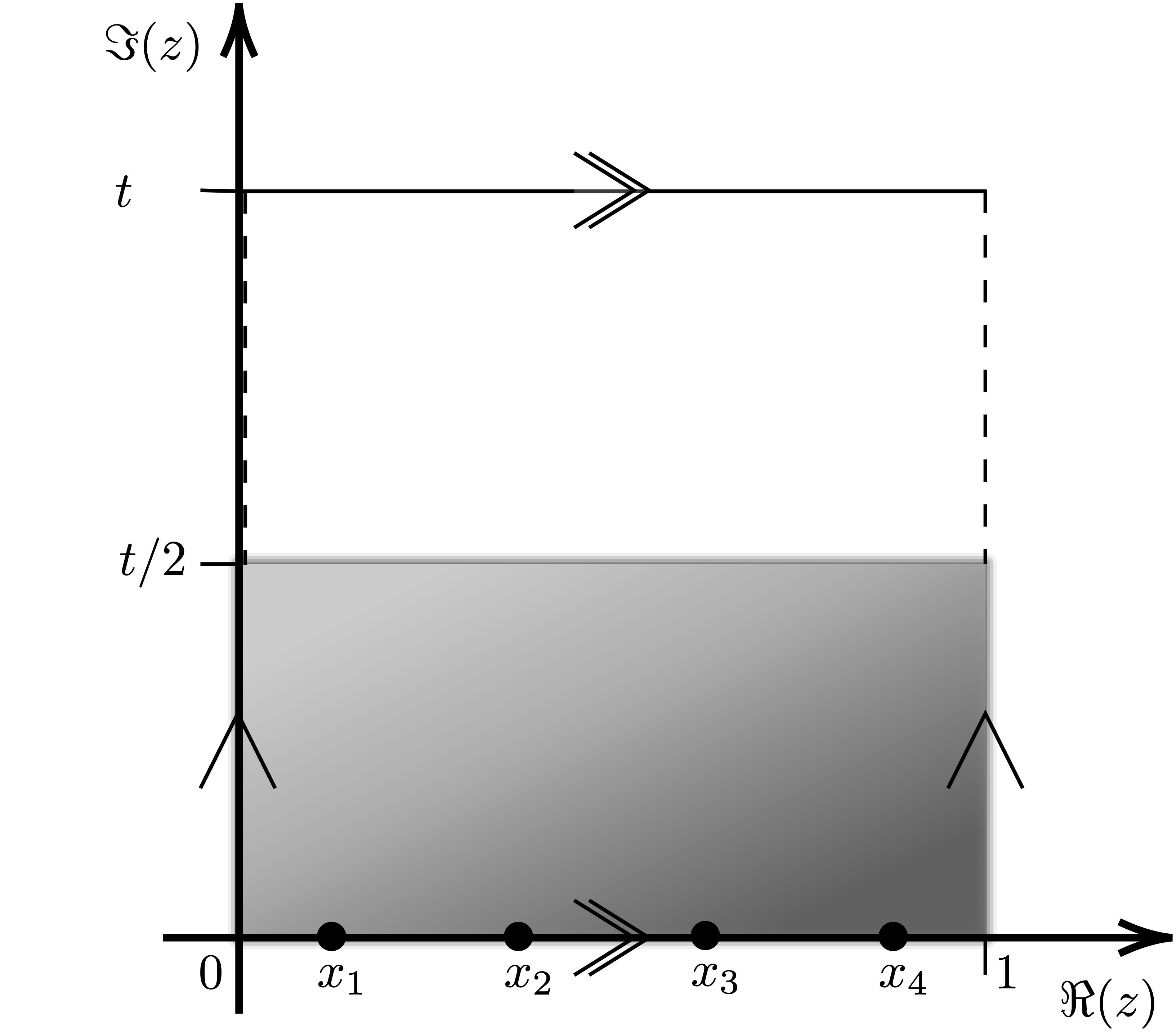}    
        }
        \caption{planar cylinder worldsheet}
        \label{fig:PlanarWorldsheetCylinder}
    \end{subfigure} \hspace{-4em}
    \begin{subfigure}[b]{0.57\textwidth}
        \resizebox{\textwidth}{!}{
        \includegraphics[width=\textwidth]{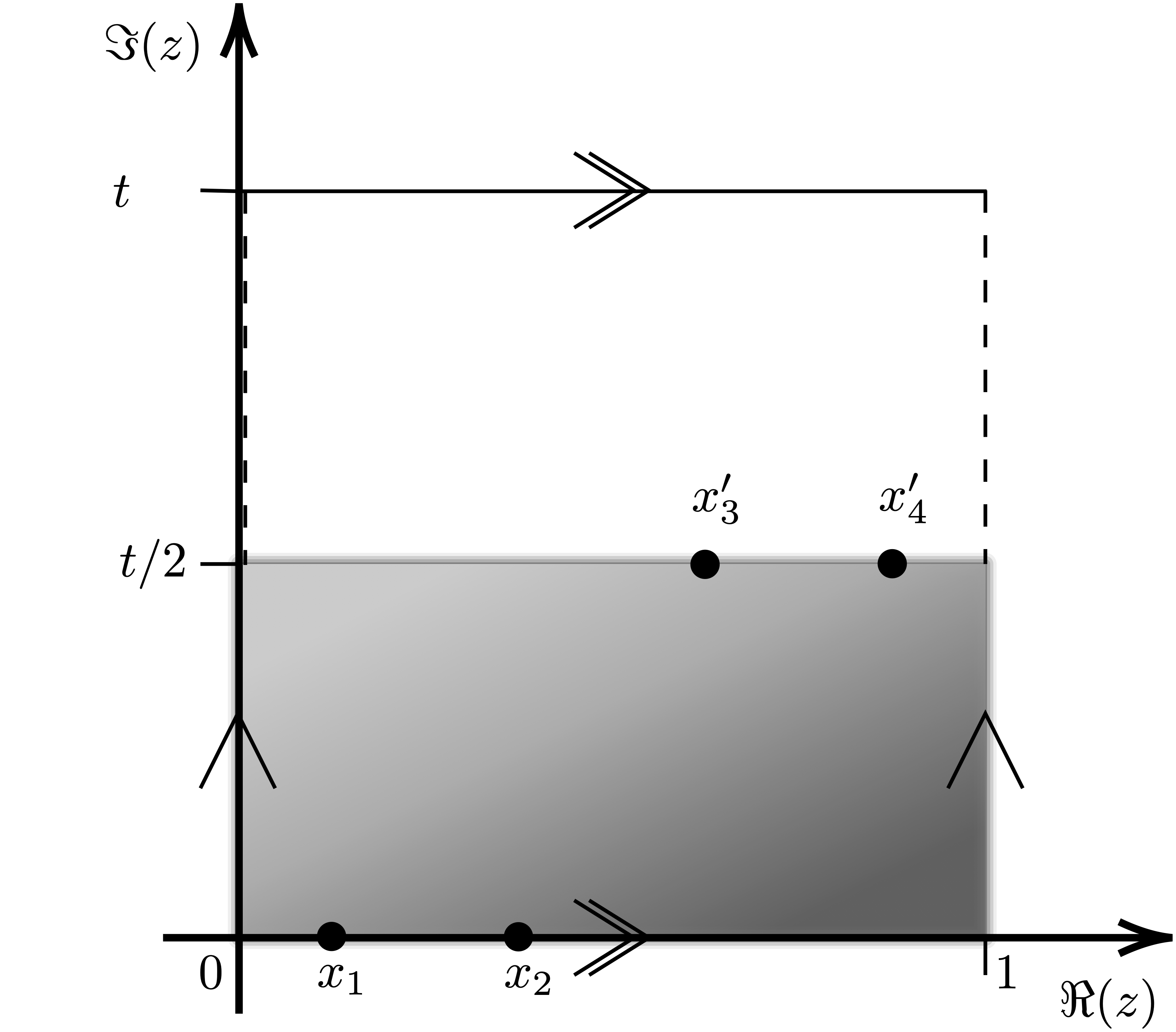}
        }
        \caption{non-planar cylinder worldsheet}
        \label{fig:NonPlanarWorldsheet}
    \end{subfigure}
    \caption[]{Orientated planar and non-planar string worldsheets with cylinder topology as seen in Figure \ref{fig:Both3DCylinders} are mapped to the complex plane depicted here by the grey shaded region. The topological identification of the edges is inherited from a covering torus with the usual modular parameter $\tau = it $ and indicated with single \tikz[x=1pt,y=1pt,yscale=-0.4,xscale=0.4]
{\draw  [color={rgb, 255:red, 0; green, 0; blue, 0 }  ,draw opacity=1 ][fill={rgb, 255:red, 0; green, 0; blue, 0 }  ,fill opacity=0 ] (317,31) -- (337,41) -- (317,51) ;} and double \tikz[x=1pt,y=1pt,yscale=-0.40,xscale=0.40]{\draw  [color={rgb, 255:red, 0; green, 0; blue, 0 }  ,draw opacity=1 ] (317,44) -- (333,54) -- (317,64)(321,44) -- (337,54) -- (321,64) ;} arrowheads.}
    \label{fig:BothCylinderWorldsheets}
\end{figure}
fully capturing the amplitude dependencies on worldsheet operator insertion positions.\footnote{Note that we suppressed the $\tau$ dependence in $P\left(z_i-z_j\right)$ for ease of notation.} Note that it only depends on position differences of operator insertions, all of which are parametrised by complex coordinates $z_{k}  = \real{(z_{k})} + i\imaginary{(z_{k})} =x_{k} + iy_{k} $. For the cylinder contributions the real part is constrained to the interval $x_{k} \in [0,1]$. The imaginary part always takes the value $y_{k}=0$ in the planar case, whereas the non-planar case also allows for $y_{k}=t/2 $. We therefore define
\begin{align}
x'_{k} = x_{k} + it/2, 
	\label{eq:OperatorInsertinWithImaginaryPart}
\end{align}
as well as $x_{ij}=x_{i}-x_{j}$ and by extension
\begin{align}
	x_{ij'} = x_{i} - x'_{j} = x_{ij} - it/2,
\end{align}
for convenience. Correspondingly, we will abbreviate the Green functions as
\begin{align}
    P_{ij} = P\left(x_{ij}\right), \qquad P_{ij'} = P\left(x_{ij'}\right) = P\left(x_{ij} - it/2\right),
    \label{eq:GreenFunctionShifted}
\end{align}
where the latter will be relevant when two operators are inserted on different boundaries. In general the Green function enters the correlators and by extension the integrands as part of an exponentiated combinatoric sum over all pairs of puncture positions accompanied by matching Mandelstam variables \cite{SCHWARZ1982223}. At four points, the external polarisations are fully captured by a kinematical prefactor 
\begin{align}
	K_{4} = s_{12}s_{23}A_{\text{YM}}^{\text{tree}}(1,2,3,4). 
\end{align}
 The full four point correlator $\mathscr{K}_{4}$ then reads      
\begin{align}
	\mathscr{K}_{4} =  s_{12}s_{23}A_{\text{YM}}^{\text{tree}}(1,2,3,4) \exp\left[\alpha' \sum_{i<j}^{4}  s_{ij}P_{ij}\right].
	\label{eq:CorrelatorFourPoint}
\end{align}
All that is left to do is setting up and computing the Integral over the string worldsheet moduli, i.e.\ all vertex operator positions $x_{1},\ldots,x_{4}$ and the now \emph{exponentiated} modular parameter $q=e^{2\pi i\tau} $. We will focus on the operator positions first.
External open string states are generally accompanied by gauge degrees of freedom encoded in Lie-algebra generators $t^{a}\in \text{SO}(N_{G})$ appearing inside trace environments in fundamental representation labelling the cyclic ordering of the vertex operators. These are known as Chan-Paton factors or simply color factors in the massless case at hand. For the planar cylinder and the M{\"o}bius strip, all operators are inserted on the same boundary allowing for a single trace $\Tr\left(t^{1}t^{2}\ldots t^{n}\right) $ only. This has important ramifications as integrating over the operator insertions has to be performed in such a way that the cyclic ordering of the positions matches the accompanying color factors. Consequently, the integration domain over the vertex operator insertions for these cases is
\begin{align}
	\mathcal{D}(1,2,\ldots,n) = \{
	 0 \leq x_{1} < x_{2} < \ldots < x_{n} < 1 , \  y_{1,2,\ldots,n}= 0 
	\}. 
	\label{eq:DomainPlanarCylinder}
\end{align}
In contrast, the non-planar cylinder by definition has operator insertions on two boundaries and thus allows for double trace factors $\Tr\left(t^{1}\ldots t^{i}\right)\Tr\left(t^{i+1}\ldots t^{n}\right) $. The corresponding integration domain reads 
\begin{align}
	\mathcal{D}(1,2,\ldots,j|j+1,\ldots,n) = \{
	0 \leq x_{1} < x_{2} < \ldots < x_{n} < 1 , \ y_{1,2,\ldots,i}= 0, \  y_{i+1,i+2,\ldots,n} = t/2 
	\}.
	\label{eq:IntegrationDomainNonPlanar} 
\end{align} 
Additionally, we can exploit the translational symmetry of genus one surfaces to set $x_{1}=0$ via a delta distribution $\delta(x_1)$ in the integrand in all cases. 
The only thing left to discuss is the integration over the exponentiated modular parameter $q= e^{2\pi i\tau} $ in each case. The planar and non-planar cylinder share the same parameter $\tau_{C} = it, \, t \in \mathbb{R}_{+}$ and by extension $q_{C} = e^{2\pi i\tau_{C}}  = e^{-2\pi t} \equiv q \in [0,1]$, resulting in an integral $\int_{0}^{1} \dd{q}/q$. For the M{\"o}bius strip we have $\tau_{M} = it + 1/2, \, t \in \mathbb{R}_{+}$ and therefore $q_{M} = e^{2\pi i\tau_{M}} = e^{2\pi i(t+1/2)} = e^{i \pi } e^{2\pi it} = - q_{C}$. The modular parameter of the M{\"o}bius strip and the cylinder are identical up to a sign. This combined with the fact that the planar cylinder and the M{\"o}bius strip possess identical integration domains over the operator punctures means that the contributions of the M{\"o}bius strip can completely be inferred from those of the planar cylinder. Accounting for all this, one can define an iterated integral encapsulating the first domain (\ref{eq:DomainPlanarCylinder}) and the translation symmetry of genus one surfaces  
\begin{align}
	\int_{1234} \equiv \int_{0}^{1} \dd{x_{4}} \int_{0}^{x_{4}} \dd{x_{3}} \int_{0}^{x_{3}} \dd{x_{2}} \int_{0}^{x_{2}} \dd{x_{1}} \, \delta(x_{1}),
	\label{eq:IntegralOverPunctures}
\end{align}
which, when combined with the $x_{i}$ dependent part of the four point string correlator (\ref{eq:CorrelatorFourPoint}) as an integrand 
\begin{align}
	I_{1234}(q) =  \int_{1234} \, \prod_{i<j}^{4} \, \exp\left[\alpha'   s_{ij}P_{ij}\right],
	\label{eq:BasicIterativeIntegral} 
\end{align}
allows us to express the planar cylinder and the M{\"o}bius strip amplitude as 
\begin{align}
	\mathcal{A}_{\text{Cyl}}\left(1,2,3,4\right) &= s_{12}s_{23}A_{\text{YM}}^{\text{tree}}(1,2,3,4) \int_{0}^{1} \frac{\dd{q}}{q} I_{1234}(q), \label{eq:CylinderContribution}  \\
	\mathcal{A}_{\text{M{\"o}b}}\left(1,2,3,4\right) &= s_{12}s_{23}A_{\text{YM}}^{\text{tree}}(1,2,3,4) \int_{0}^{1} \frac{\dd{q}}{q} I_{1234}(-q).
	\label{eq:MoebiusContribution}  
\end{align}
A priori, one would expect both terms to enter the total amplitude with a factor of $N_{G}$ from the shared group theoretic trace factor $\Tr(t^{1}t^{2}t^{3}t^{4})$. However, another subtlety worth addressing when combining these two terms is the need to account for the single boundary of the M{\"o}bius strip being twice as long any one of the two separate cylinder boundaries (c.f. subfigures \ref{fig:PlanarWorldsheetCylinder} and \ref{fig:WorldsheetMoebiusStrip}) by inserting a factor of $-32/N_{G}$ for the M{\"o}bius strip \cite{greenInfinityCancellations321985}. For operators inserted on two different boundaries as depicted in figure \ref{fig:NonPlanarWorldsheet}, we define 
\begin{align}
	I_{12|34}(q) = \left(\int_{0}^{1} \prod_{i=1}^{4} \dd{x_{i}}\right) \delta(x_{1}) \exp\Biggr[\alpha'\biggl(s_{12}P_{12} + s_{34}P_{34} + \sum_{\substack{i=1,2 \\ 
    j=3,4}}  s_{ij}P_{ij'} \biggr)\Biggr].
	\label{eq:NonPlanarPunctureIntegral}
\end{align}
Note that in (\ref{eq:NonPlanarPunctureIntegral}) the integrals over the individual $x_{i}$'s are all performed over the same domain $[0,1]$ and not iterated anymore like (\ref{eq:IntegrationDomainNonPlanar}) would suggest. This can be achieved by using the symmetry of the accompanying color factors. Their tracelessness is also the the reason why we do not need to write down an integral for three punctures on one side and one on the other, i.e. something like $I_{123|4}$, as the associated color factors $\Tr(t^{1}t^{2}t^{3})\Tr(t^{4})$ simply vanish. The non-planar cylinder amplitude contributions read
\begin{align}
	\mathcal{A}_{\text{NP}} = s_{12}s_{23}A_{\text{YM}}^{\text{tree}}(1,2,3,4) \int_{0}^{1} \frac{\dd{q}}{q} I_{12|34}(q).
\end{align}        
Gathering all the terms, inserting the previously suppressed color factors, adding appropriate powers of the ten dimensional Yang-Mills coupling constant $g_{10}$ combined with $\alpha'$ in line with \cite{greenSuperstringTheory2012} and accounting for cyclic symmetry, the full one loop massless open superstring amplitude reads 
\begin{align}
	\mathcal{A}_{1,4}^{\text{String}} &= \frac{g_{10}^{2}}{\alpha'}s_{12}s_{23}A_{\text{YM}}^{\text{tree}}(1,2,3,4) \int_{0}^{1} \frac{\dd{q}}{q} \Bigl\{
	\Tr(t^{1}t^{2}t^{3}t^{4})\left[
	N_{G} I_{1234}(q)-32 I_{1234}(-q)
	\right] \nonumber \\ 
	&\phantom{=}+ \Tr(t^{1}t^{2})\Tr(t^{3}t^{4})I_{12|34}(q) + \text{cyc}\left(2,3,4\right)
	\Bigr\}.
	\label{eq:OneLoopStringAmplitudeFull} 
\end{align}
As previously stated, in \cite{broedelEllipticMultipleZeta2015,broedelTwistedEllipticMultiple2018b} the authors managed to compute this remarkably difficult integral in terms of an $\alpha'$-expansion. This essentially amounts to Taylor expanding the exponential in the integrals $I_{1234}(q)$ and $I_{12|34}(q)$ and performing the integration over the string punctures $x_{i}$ to ever increasing order in $\alpha'$. For all elaborate details pertaining to this computation we refer to these papers. Here we will just state the results. In the planar sector of the amplitude, one can show
\begin{align}
 I_{1234}\left(q\right) &= \varphi(0,0,0) \, -  \left(\alpha'\right) \, 2\, \varphi(0,1,0,0) \, (s_{12}+s_{23})  \, \nonumber  \\ 
  &\phantom{=} + \,  \left(\alpha'\right)^2 \left[ \, 2 \, \varphi(0,1,1,0,0)\,  \big( s_{12}^2  + s_{23}^2 \big)    
 - \, 2 \,  \varphi(0,1,0,1,0) \,s_{12}s_{23} \,\right]  \nonumber  \\
 &\phantom{=} + \, \left(\alpha'\right)^3 \left[ \sigma_5 \, (s_{12}^3+2 s_{12}^2 s_{23} + 2s_{12} s_{23}^2+s_{23}^3)
\, + \, \sigma_{2,3} \, s_{12} s_{23}(s_{12}+s_{23})\right]  
+  {\cal O}(\alpha'^4),
\label{eq:PlanarPunctureIntegral}
\end{align}
with 
\begin{align}
\sigma_5 &=  \frac{4}{3} \, \big[ \varphi(0,0,1,0,0,2)+\varphi(0,1,1,0,1,0) - \varphi(2,0,1,0,0,0) - \zeta_2\, \varphi(0,1,0,0) \big] \label{eq:emzvCoefficients1} \\
\sigma_{2,3} &= - \frac{1}{3} \varphi(0, 0, 1, 0, 2, 0)+ 
 \frac{3}{2} \varphi(0, 1, 0, 0, 0, 2) +
 \frac{1}{2} \varphi(0, 1, 1, 1, 0, 0) \nonumber \\
 & \ \ \ +
 2 \varphi(2, 0, 1, 0, 0, 0) + 
 \frac{4}{3} \varphi(0, 0, 1, 0, 0, 2) +
 \frac{10}{3} \zeta_2 \, \varphi(0, 1, 0, 0) \ .
 \label{eq:emzvCoefficients2}
\end{align}
Here $\varphi(0,0,0)$, $\varphi(0,1,0,0)$ etc. denote the previously alluded to eMZVs.\footnote{In the source papers \cite{broedelEllipticMultipleZeta2015,broedelTwistedEllipticMultiple2018b}, the eMZVs and the sums thereof in (\ref{eq:emzvCoefficients1}) and (\ref{eq:emzvCoefficients2}) were labelled with $\omega(\bullet,\ldots,\bullet)$ and $\beta_{\bullet}$ instead of $\varphi(\bullet,\ldots,\bullet)$ and $\sigma_{\bullet}$ respectively. We did so in order to avoid a clash with common celestial holography notation where $\omega$ as seen in (\ref{eq:FourMomenta}) denotes the energy of a scattered particle and $\beta$ denotes the sum over the imaginary part of the conformal scaling dimensions (\ref{eq:beta}).} They generically take the form of double infinite series over positive integers $(n,m) \in \mathbb{N}_{\geq 1}$ involving the exponentiated modular parameter $q$. We have for instance     
\begin{align}
    & \varphi(0,0,0) = \frac{1}{6}, \\ 
    & \varphi(0,1,0,0) = \frac{\zeta_{3}}{8\zeta_{2}} + \frac{3}{2\pi^2}\sum_{m,n\geq 1}^{\infty}  \frac{1}{m^3} q^{mn}.
\end{align}
Note that for example at order $(\alpha')^{0}$, we have for \emph{both} the planar cylinder and the M{\"o}bius strip the remaining integral 
\begin{align}
	\int_{0}^{1} \frac{\dd{q}}{q} \varphi(0,0,0) = \frac{1}{6} \left.\log(q)\right|_{0}^{1},
	\label{eq:eMZVDivergentExample1} 
\end{align}
which diverges. Equation (\ref{eq:OneLoopStringAmplitudeFull}) then famously forces us to choose $N_{G}=32$ in order for the two divergent pieces from the planar cylinder and the M{\"o}bius strip to cancel. For more  examples of eMZVs we refer the reader to Appendix \ref{sec:eMZVs}. Their regularisation will be dealt with  in detail in chapter \ref{sec:PlanarLoopLevelCelestialExpansion}. The important takeaway for the celestial minded person here, is that eMZVs are independent of the energy $\omega$ and therefore merely bystanders in the Mellin transform. Of course the opposite is true for every Mandelstam variable since according to according to (\ref{eq:MandelstamVariable}) we have $s_{ij} \propto \omega^2$. For the non-planar sector one finds
\begin{multline}
I_{12|34}(q) = q^{\alpha' s_{12}/4} \bigg\{ 1+ s_{12}^2  \left( \frac{ 7\zeta_2 }{6} + 2\,\varphi(0,0,2)  \right)
+2s_{23}\left(s_{12}+s_{23}\right) \left( \frac{ \zeta_2 }{3} + \varphi(0,0,2)  \right)  \\
  - 4 \, \zeta_2 \, \varphi(0,1,0,0) \, s_{12}^3 - s_{12}s_{23}\left(s_{12}+s_{23}\right) \left(
 \frac{5}{3}\,\varphi(0,3,0,0) + 4 \, \zeta_2 \, \varphi(0,1,0,0)  - \frac{1}{2} \zeta_3
 \right) + \mathcal{O}(\alpha'^4) \bigg\},	
\end{multline}
Note that the overall structure between this and the planar case (\ref{eq:PlanarPunctureIntegral}) is very similar except for the presence of the global factor $q^{\alpha' s_{12}/4}$.

\section{Celestial String Tree Amplitudes}
\label{sec:CelestialTrees}

In this section we give a lightening review of relevant results for celestial string tree amplitudes found in \cite{stiebergerStringsCelestialSphere2018}. They will mainly serve as a frame of reference for results to come.

Taking the aforementioned four point open string type \MakeUppercase{\romannumeral 1} amplitude (\ref{eq:StringTreeAmplitude}) and computing its celestial counterpart via (\ref{eq:CelestialAmplitude}), one eventually obtains the closed form expression
\begin{align}
    \tilde{\mathcal{A}}_{\text{String}}^{\text{tree}}(1^-,2^-,3^+,4^+) = 4 \, g_{10}^2 \left(\alpha'\right)^\beta \delta(r-\bar{r})\theta(r-1) K\left(h_i,\bar{h}_i\right) r^{\frac{5-\beta}{3}}\left(r-1\right)^{\frac{2-\beta}{3}} I(r,\beta).
    \label{eq:CelestialStringTree}
\end{align}
The term $I(r,\beta)$ denotes the last remaining non trivial energy integral and is defined as
\begin{align}
    I(r,\beta) = \pi\delta(i\beta) + \frac{1}{2}\Gamma\left(\beta\right)\Gamma\left(1-\beta\right)\left(-r\right)^{\beta} \sum_{k=1}^{\infty} \left(-r\right)^{-k}S(\beta-k,\beta),
    \label{eq:TreeEnergyIntegral}
\end{align}
where $S(\beta-k,\beta)\equiv S_{\beta-k,k}(1)$ denotes a special case of the \emph{Nielsen's polylogarithm}
\begin{align}
	S_{n,k}\left(t\right) = \frac{(-1)^{n+k-1}}{\Gamma(\beta-k)\Gamma(k+1)} \int_{0}^{1} \frac{\dd{x}}{x} \ln^{n-1} \left(x\right)\ln^{k} \left(1-xt\right), \quad t\in\mathbb{C}.
\end{align}
In a stunning display of consistency, these Nielsen polylogarithms can actually be expressed as the aforementioned MZVs (\ref{eq:GeneralMZVOfDepthK}), specifically at depth $k$ as $S_{n,k}(1) = \zeta\left(n+1,\{1\}^{k-1}\right) $.\footnote{Here one denotes $\{1\}^{k-1}=\underbrace{1,\ldots,1}_{(k-1)\text{-times}}$.} Consequently, after performing an index shift in $k$ and employing Euler's reflection formula $\Gamma(\beta)\Gamma(1-\beta) = \pi/\sin{\pi\beta}$, one can show  
\begin{align}
	I(r,\beta) = \pi\delta(i\beta) + \frac{1}{2}\frac{\pi}{\sin{\pi\beta}} \left(-r\right)^{\beta-1} \sum_{k=0}^{\infty} \left(-r\right)^{-k}\zeta\left(\beta-k,\{1\}^{k}\right).
	\label{eq:TreeEnergyIntegral2}
\end{align}
A striking feature of this result is that the entire $\alpha'$-dependence is fully captured by a simple prefactor $(\alpha')^{\beta}$. Evaluating (\ref{eq:CelestialStringTree}) on the support of $\delta(i\beta)$ present in the first term of  (\ref{eq:TreeEnergyIntegral2}) makes all $\alpha'$ dependency drop out and consistently reproduces the YM gluon amplitude \cite{pasterskiGluonAmplitudes2d2017} as expected from the  full string amplitude. By contrast, the higher terms do not allow for such manipulations. The infinite series in (\ref{eq:TreeEnergyIntegral2}) represents the commonly found stringy tower of massive states.
\section{Celestial String Loop Integrands}
\label{sec:CelestialLoopIntegrands}

Now that we have assembled all necessary ingredients, there are two routes of celestialization we would like present. The first one, discussed in this chapter, involves taking the fully non-integrated string amplitude integrand as seen in (\ref{eq:OneLoopStringAmplitudeFull}) and performing a Mellin transform with respect to the energy of all external string states.\footnote{For full disclosure we want to mention that while writing up the very early results of this paper, including the one in (\ref{eq:CelestialCylinderIntegrand}), a similar result appeared in \cite{donnayCelestialOpenStrings2023a}. We encourage readers to consult their work for further insights in celestial string amplitudes, in particular how to compute the field theory limit.}  We dub the resulting expression celestial string integrands, emphasising the remaining integrals over the string punctures $x_{i}$ and the moduli parameter $q$.

\subsection{Celestial Planar Cylinder \& M{\"o}bius Strip Integrand}

The computations of the celestial planar cylinder and M{\"o}bius strip integrand are virtually identical, so we will focus exclusively on the cylinder first. Suppressing the $N_{G}$ constant and the color factors, we define the string integrand for the cylinder as 
\begin{align}
\mathscr{I}_{\text{Cyl}}\left(q,\{x_{i}\}\right) = \frac{g_{10}^2}{\alpha'} s_{12}s_{23} \, \mathcal{A}_{\text{YM}}^{\text{tree}}\left(1^{-},2^{-},3^{+},4^{+}\right) \prod_{i<j}^{4} \exp\left[\alpha' s_{ij}P_{ij}\right].  
\end{align}
For the purpose of the Mellin transform, we need to account for all terms that are dependent on the energies of the four scattered open strings $\omega_{i}, i\in(1,\ldots,4)$. Starting with the product of exponentials, using (\ref{eq:MandelstamEquivalents}), (\ref{eq:MandelstamMasslessCondition}) to simplify the terms, factoring out an $s_{12}$, expressing the resulting ratio of Mandelstams as the conformal cross ratio (\ref{eq:ConformalCrossRatio}) and expressing the remaining $s_{12}$ explicitly in terms of celestial sphere coordinates (\ref{eq:MandelstamVariable}), we obtain 
\begin{align}
    \prod_{i<j}^{4} \exp\left[\alpha' s_{ij}P_{ij}\right]  
    =  \exp\left[-\alpha' \omega_{1}\omega_{2} z_{12}\bar{z}_{12} \left(\mathcal{P}_{1234} +  \frac{\mathcal{P}_{1423}}{r} \right) \right],
\end{align}
where we introduced
\begin{align}
    \mathcal{P}_{1234} =  P_{13} + P_{24} - P_{12} - P_{34}, \qquad \mathcal{P}_{1423} =  P_{13} + P_{24} - P_{14} - P_{23},
    \label{eq:GreenFunctionCombinationPlanar} 
\end{align}
which do not depend on any $\omega_{i}$ and can therefore be treated as constants when performing the Mellin transform. The other terms are simply Mandelstam variables and the tree-level Yang-Mills gluon amplitude $\mathcal{A}_{\text{YM}}^{\text{tree}}$ given by (\ref{eq:MandelstamVariable}) and (\ref{eq:ParkeTaylor}) respectively. Putting everything together, the celestial cylinder string integrand can be computed as
\begin{align}
	\Tilde{\mathscr{I}}_{\text{Cyl}}\left(q,\{x_{i}\}\right) &= \prod_{i=1}^{4} \int_{0}^{\infty} \dd{\omega_{i}} \omega_{i}^{\Delta_{i}-1} \, \mathscr{I}_{\text{Cyl}}\left(q,\{x_{i}\}\right) \nonumber \\ 
	&=  \frac{4\, g_{10}^2}{\alpha'}\, \delta(r-\bar{r}) \theta\left(r-1\right) \frac{z_{12}\bar{z}_{12} }{z_{14}\bar{z}_{14}} \frac{z_{12}^3}{z_{23}z_{34}z_{41}}   \prod_{i=1}^{4}  \int_{0}^{\infty} \dd{\omega_{i}} \prod_{j=1}^{3} \delta\left(\omega_j - \chi_{j} \omega_{4}\right) \nonumber \\ 
	&\phantom{=} \times \omega_{1}^{\Delta_{1}+1}\omega_{2}^{\Delta_{2}+2}\omega_{3}^{\Delta_{3}-1}\omega_{4}^{\Delta_{4}-3} \exp\left[-\alpha' \omega_{1}\omega_{2} z_{12}\bar{z}_{12} \left(\mathcal{P}_{1234} +  \frac{\mathcal{P}_{1423}}{r} \right) \right] \nonumber \\ 
	&= \frac{4 \, g_{10}^2}{\alpha'} \delta(r-\bar{r}) \theta\left(r-1\right) \frac{z_{12}\bar{z}_{12} }{z_{14}\bar{z}_{14}} \frac{z_{12}^3}{z_{23}z_{34}z_{41}} \chi_{1}^{\Delta_{1}+1}\chi_{2}^{\Delta_{2}+2}\chi_{3}^{\Delta_{3}-1} \nonumber \\ 
    &\phantom{=} \times \int_{0}^{\infty} \dd{\omega_{4}} \omega_{4}^{\Delta-1} \exp\left[-\alpha' \chi_{1}\chi_{2}  z_{12}\bar{z}_{12} \left(\mathcal{P}_{1234} +  \frac{\mathcal{P}_{1423}}{r} \right) \omega_{4}^2 \right] \nonumber \\ 
    &= \frac{2\, g_{10}^2}{\alpha'}\, \delta(r-\bar{r}) \theta\left(r-1\right) \frac{\left(z_{12}\bar{z}_{12}\right)^{1-\frac{\Delta}{2}} }{z_{14}\bar{z}_{14}} \frac{z_{12}^3}{z_{23}z_{34}z_{41}} \chi_{1}^{\Delta_{1}+1-\frac{\Delta}{2}}\chi_{2}^{\Delta_{2}+2-\frac{\Delta}{2}}\chi_{3}^{\Delta_{3}-1} \nonumber \\ 
    &\phantom{=} \times \left(\alpha' \left(\mathcal{P}_{1234} +  \frac{\mathcal{P}_{1423}}{r} \right)\right)^{-\frac{\Delta}{2}}\Gamma\left(\frac{\Delta}{2} \right) \nonumber \\ 
&=  2 \, g_{10}^2\, \left(\alpha'\right)^{\beta-3}\Gamma\left(2-\beta \right)   K\left(h_i,\bar{h}_i\right)  \delta(r-\bar{r}) \theta\left(r-1\right)  r^{\frac{5-\beta}{3}}(r-1)^{\frac{2-\beta}{3}} \nonumber \\
 &\phantom{=} \times r^{\beta-1} \left(\mathcal{P}_{1234} +  \frac{\mathcal{P}_{1423}}{r} \right)^{\beta-2}.
 \label{eq:CelestialCylinderIntegrand}
\end{align}  
These steps warrant some explanation. From the second to the third equals sign we evaluated all but the $\omega_{4}$ integral on the support of the delta distributions arising from the momentum conserving delta distributions (\ref{eq:MomentumDeltaCelestialCoordinates}). The transition from the third to fourth equal signs is accomplished via the integral representation of the $\Gamma$-function. In the last step we performed algebraic manipulations, making generous use the constraints implied by $\delta(r-\bar{r}) $ and the definition of $\beta$ (\ref{eq:beta}) to show  
\begin{align}
    \frac{\left(z_{12}\bar{z}_{12}\right)^{1-\frac{\Delta}{2}} }{z_{14}\bar{z}_{14}} \frac{z_{12}^3}{z_{23}z_{34}z_{41}} \chi_{1}^{\Delta_{1}+1-\frac{\Delta}{2}}\chi_{2}^{\Delta_{2}+2-\frac{\Delta}{2}}\chi_{3}^{\Delta_{3}-1} = K\left(h_i,\bar{h}_i\right) r^{\frac{5-\beta}{3}}(r-1)^{\frac{2-\beta}{3}}r^{\beta-1}.   
\end{align}
Furthermore, we dropped an overall constant phase factor $(-1)^{\frac{4}{3}(1+\beta)+i(\lambda_2+\lambda_3)}$ for simplicity. As expected from conformal symmetry, the collection of all factors of celestial coordinates $z_{ij}$ can be absorbed into a product of the covariant prefactor $K\left(h_i,\bar{h}_i\right)$ and powers of the invariant conformal cross ratios $r$. A remarkable feature of the result in (\ref{eq:CelestialCylinderIntegrand}) is that the $\alpha'$ dependence of the amplitude is reduced to a mere global factor $(\alpha')^{\beta-3}$. The same phenomenon occurred at tree level as seen in section \ref{sec:CelestialTrees}, albeit with a slightly shifted power $(\alpha')^{\beta}$. 

As seen in (\ref{eq:OneLoopStringAmplitudeFull}), the M{\"o}bius amplitude can be obtained from those of the planar cylinder by the simple replacement $q \to -q$ and and multiplying by an overall factor $-32/N_{G}$. In particular, this does change anything with regards to the Mellin transform apart from bookkeeping and so the celestial Integrand of the M{\"o}bius strip is simply 
\begin{align}
	\tilde{\mathscr{I}}_{\text{M{\"o}b}}\left(q,\{x_{i}\}\right) = -\frac{32}{N_{G}} \tilde{\mathscr{I}}_{\text{Cyl}}\left(-q,\{x_{i}\}\right).
	\label{eq:CelestialMoebiusIntegrand}
\end{align}

\subsection{Celestial Non-Planar Cylinder Integrand}

The celestial integrand for the non-planar cylinder amplitude differs from its planar cousin in two aspects. The first one being the double trace color factors $\Tr\left(t^{1}t^{2}\right)\Tr\left(t^{3}t^{4}\right)$ and cyclically distinct permutations thereof, which we will suppress like before. The second one is the occurrence of genus-one Green function with arguments shifted by $-it/2$ as seen in (\ref{eq:GreenFunctionShifted}) due to insertion of the operators on both boundaries of the worldsheet. Rewriting the exponents in the non-planar puncture integral (\ref{eq:NonPlanarPunctureIntegral}) using Mandelstam properties (\ref{eq:MandelstamEquivalents}) and (\ref{eq:MandelstamMasslessCondition}) like before, we compute       
\begin{align}
	\exp\Biggr[\alpha'\biggl(s_{12}P_{12} + s_{34}P_{34} + \sum_{\substack{i=1,2 \\ 
    j=3,4}}  s_{ij}P_{ij'} \biggr)\Biggr] = \exp\left[-\alpha' \omega_{1}\omega_{2} z_{12}\bar{z}_{12} \left(\mathcal{P}'_{1234} +  \frac{\mathcal{P}'_{1423}}{r} \right) \right],
\end{align}
where we have defined
\begin{align}
    \mathcal{P}'_{1234} =  P_{13'} + P_{24'} - P_{12} - P_{34}, \qquad \mathcal{P}'_{1423} =  P_{13'} + P_{24'} - P_{14'} - P_{23'},  
\end{align}
in complete analogy to (\ref{eq:GreenFunctionCombinationPlanar}) but adding a tilde on every propagator with pairwise operator insertions on different boundaries, i.e. for all cases except $P_{12}$ and $P_{34}$ in accordance with the chosen fixed color structure. 
Since all other relevant aspects for the Mellin transform are equal to the planar cylinder case, we can infer the non-planar cylinder contribution from it by a suitable replacement $(\mathcal{P}_{1234},\mathcal{P}_{1423}) \to (\mathcal{P}'_{1234},\mathcal{P}'_{1423})$
\begin{align}
\Tilde{\mathscr{I}}_{\text{NP}}\left(q,\{x_{i}\}\right) &= 2 \, g_{10}^2\, \left(\alpha'\right)^{\beta-3}\Gamma\left(2-\beta \right)   K\left(h_i,\bar{h}_i\right)  \delta(r-\bar{r}) \theta\left(r-1\right)  r^{\frac{5-\beta}{3}} (r-1)^{\frac{2-\beta}{3}} \nonumber \\
 &\phantom{=}\times r^{\beta-1} \left(\mathcal{P}'_{1234} +  \frac{\mathcal{P}'_{1423}}{r} \right)^{\beta-2}.
 \label{eq:CelestialNonPlanarIntegrand}	
\end{align}

\subsection{Remaining String Moduli Integrals}
\label{sec:RemainingStringModuliIntegrals}

This subsection explores the next steps necessary if one would want to obtain the fully fledged celestial string one loop amplitude starting from the results of this section. In doing so, we aim to provide a starting point for future research endeavours, as well as highlighting the inherent difficulties in this approach which made solving this problem intractable, at least in the context of this paper.

In order to obtain the fully integrated celestial string amplitude at one loop, one needs to perform the remaining integrals over the string moduli, i.e. the worldsheet insertion points $x_{i}$ and the modular parameter $q$. In the following we will focus on the planar cylinder contribution for simplicity, but this discussion can easily be transferred to the M{\"o}bius strip and non-planar sector via the aforementioned changes displayed in (\ref{eq:CelestialMoebiusIntegrand}) and (\ref{eq:CelestialNonPlanarIntegrand}) respectively. The computation amounts to 
\begin{align}
	\Tilde{\mathcal{A}}_{_{\text{Cyl}}}\left(1,2,3,4\right) &=  \int_{0}^{1} \frac{\dd{q}}{q} \, \int_{1234} \, \Tilde{\mathscr{I}}_{\text{Cyl}}\left(q,\{x_{i}\}\right) \nonumber \\ 
	&= 2 \, g_{10}^2\, \left(\alpha'\right)^{\beta-3}\Gamma\left(2-\beta \right)   K\left(h_i,\bar{h}_i\right)  \delta(r-\bar{r}) \theta\left(r-1\right)  r^{\frac{5-\beta}{3}}(r-1)^{\frac{2-\beta}{3}} \nonumber \\
 &\phantom{=} \times r^{\beta-1}  \int_{0}^{1} \frac{\dd{q}}{q} \, \int_{1234} \, \left(\mathcal{P}_{1234} +  \frac{\mathcal{P}_{1423}}{r} \right)^{\beta-2}.
 \label{eq:AnsatzRemainingIntegralAfterMellin}
\end{align}
It is at this point that one needs to specify what the combinations of genus one Green functions $\mathcal{P}_{1234}$ and $\mathcal{P}_{1423}$ look like precisely. For operator insertions along the real line of the worldsheet, as is the case for the planar cylinder amplitude, it suffices to consider a piece of the holomorphic part of the $q$-Expansion of the \emph{Eisentein-Kronecker} series \cite{broedelEllipticMultipleZeta2015}. It reads
\begin{align}
g^{(1)}\left(z\right) = \pi\cot{(\pi z)} + 4\pi\sum_{m=1}^{\infty}\sin{(2\pi mz)}\sum_{n=1}^{\infty} q^{mn}.	
\end{align}
The genus one Green function $P_{ij}$ can then be defined as an integral of this function 
\begin{align}
	P_{ij} = \int_{0}^{x_{ij}} \dd{u} g^{(1)}\left(u\right) = \lim_{\varepsilon\to 0}\log\left[\frac{\sin{\pi x_{ij}}}{\sin{\pi \varepsilon}}\right] + 4 \sum_{m=1}^{\infty}\sin^2{(\pi m x_{ij})}\sum_{n=1}^{\infty} \frac{q^{mn}}{m},
	\label{eq:GenusOneGreenExplicit} 
\end{align}
where one needs to introduce a regularisation parameter $\varepsilon>0$ to avoid the singularity from the logarithm at $u =0$. Computing $\mathcal{P}_{1234}$ then amounts to using its definition (\ref{eq:GreenFunctionCombinationPlanar}) and inserting the previous result (\ref{eq:GenusOneGreenExplicit}). We ultimately obtain 
\begin{align}
	\mathcal{P}_{1234} &= 4 \sum_{m=1}^{\infty}\left(\sin^2{(\pi m x_{13})} + \sin^2{(\pi m x_{24})} - \sin^2{(\pi m x_{12})} - \sin^2{(\pi m x_{34})} \right)\sum_{n=1}^{\infty} \frac{q^{mn}}{m}  \nonumber \\ 
	&\phantom{=} + \log\left[\frac{\sin{\left(\pi x_{13}\right)}\sin{\left(\pi x_{24}\right)}}{\sin{\left(\pi x_{12}\right)}\sin{\left(\pi x_{34}\right)}}\right].
	\label{eq:GenusOneGreenCombinationP1234}
\end{align}
And with a completely analogous computation
\begin{align}
	\mathcal{P}_{1423} &= 4 \sum_{m=1}^{\infty}\left(\sin^2{(\pi m x_{13})} + \sin^2{(\pi m x_{24})} - \sin^2{(\pi m x_{14})} - \sin^2{(\pi m x_{23})} \right)\sum_{n=1}^{\infty} \frac{q^{mn}}{m}  \nonumber \\ 
	&\phantom{=} + \log\left[\frac{\sin{\left(\pi x_{13}\right)}\sin{\left(\pi x_{24}\right)}}{\sin{\left(\pi x_{14}\right)}\sin{\left(\pi x_{23}\right)}}\right].
	\label{eq:GenusOneGreenCombinationP1423}
\end{align}
 Note that the divergent $\log\left[\sin\left(\pi \varepsilon\right)\right]$ terms cancelled one another, making the final result independent of the regularisation parameter. Looking at the full expressions for $\mathcal{P}_{1234}$ and $\mathcal{P}_{1423}$, it quickly becomes apparent that solving the remaining moduli space integral 
 \begin{align}
 	\Psi_{\text{cyl}}\left(\beta\right) = \int_{0}^{1} \frac{\dd{q}}{q} \, \int_{1234} \, \left(\mathcal{P}_{1234} +  \frac{\mathcal{P}_{1423}}{r} \right)^{\beta-2},
 	\label{eq:RemainingIntegralBeta}
 \end{align} 
for a generic complex parameter $\beta \in \mathbb{C}$ is a challenging task for all but the simplest (analytically continued) integer values of $\beta$. We will compute and comment on a few of these special cases in section \ref{sec:RemainingIntegralEvaluated}. However, for the general case there are an infinite amount of terms within the bracket raised to some complex power. It might be helpful disentangling the terms using the multinomial theorem for complex numbers or its contour-integral equivalent, the  Mellin-Barnes representation  \cite{dubovykMellinBarnesIntegralsPrimer2022}. Ultimately one would end up with all possible products of the genus one Green function, each individually raised to some (possibly complex) power. Some of the simpler resulting integrals would read, for example 
\begin{align}
	\int_{1234} \, \left(P_{12}\right)^{k}\left(\frac{P_{34}}{r}\right)^{\left(\beta-2\right)-k}, \qquad \int_{1234} \, \left(\frac{P_{12}}{r}\right)^{\left(\beta-2\right)}, \qquad k \in \mathbb{N}_{\geq 0},
	\label{eq:IntegralsGreenFunctionCelestial}
\end{align}
and so on. Interestingly and perhaps somewhat unsurprisingly, these terms look very similar to the ones computed in the $\alpha'$-expansion in chapter 4.2 in  \cite{broedelEllipticMultipleZeta2015}, the only computationally meaningful difference being that the celestial computation also allows for complex and negative exponents of the Green functions. Continuing along this line of thought and systematically solving all integrals of the type (\ref{eq:IntegralsGreenFunctionCelestial}) promises to be a fruitful endeavour. We leave this for future work.   

\section{Expansion of Celestial Loops Amplitudes}
\label{sec:ExpansionCelestialLoopAmplitudes}

As seen in the previous chapter, computing the full celestial string one loop amplitude by performing the string moduli space integrals over the celestial integrands amounts to an exceedingly difficult computation. In this section, we employ an alternative approach by first using the known $\alpha'$-expansions of the amplitude \cite{broedelEllipticMultipleZeta2015,broedelTwistedEllipticMultiple2018b} as summarised in subsection \ref{sec:StringLoopAmplitudes} before applying the Mellin transform. A fundamental limitation of this truncated low-energy expansion approach is that it unavoidably does not account for the high-energy contributions to the amplitude, whereas the Mellin transform places both low- and high-energy contributions on the same footing. Nevertheless, some interesting structure arises in this approach making it a  worthwhile endeavour.

Motivated by both inherent interest in the result itself and in order to develop some intuition for this approach which will be useful in the more difficult case at one loop, we will first consider celestial string amplitudes at tree level. This has the distinct advantage that at tree level, the fully integrated celestial amplitude as seen in section \ref{sec:CelestialTrees} is already known, allowing for an explicit comparison between both results. All of these considerations are gathered in subsection \ref{sec:TreeLevelCelestialExpansion}. The same ansatz is then employed for the planar and non-planar string one loop contributions in \ref{sec:PlanarLoopLevelCelestialExpansion} and \ref{sec:NonPlanarLoopLevelCelestialExpansion} respectively. Furthermore, the additional moduli space integrals $\int \dd{q}$ are evaluated for all cases at loop level using an $\varepsilon$-regularisation method. Finally, in section \ref{sec:RemainingIntegralEvaluated} we establish a tentative connection between the results obtained in this and the previous section for the loop level planar cylinder contribution.  
\subsection{Tree-level amplitudes}
\label{sec:TreeLevelCelestialExpansion}

Our ansatz consists of utilising the $\alpha'$-expansion of the string form factor (\ref{eq:ExpansionVeneziano}) and focusing exclusively on $\omega$ dependent terms relevant for the Mellin transform. In this way, we can express every part in the expansion as a product of the original YM amplitude and Mandelstam variables raised to some non-negative integer powers $n,m \in \mathbb{N}_{\geq 0}$, that is  
\begin{align}
	\mathcal{A}_{\text{YM}}^{\text{tree}}(1^-,2^-,3^+,4^+)(\alpha' s_{12})^n(\alpha' s_{23})^m,
\end{align}
the Mellin transform of which can be performed as 
\begin{align}
      \mathscr{J}_{\text{tree}}(n,m) &\equiv  g_{10}^2 \, \prod_{i=1}^{4} \int_{0}^{\infty} \dd{\omega_{{i}}}^{\Delta_i -1} \mathcal{A}_{\text{YM}}^{\text{tree}}(1^-,2^-,3^+,4^+)(\alpha' s_{12})^n(\alpha' s_{23})^m \nonumber \\ 
     &= 4\, g_{10}^2 (\alpha')^{n+m} \frac{(z_{12}\bar{z}_{12})^{n}(z_{23}\bar{z}_{23})^{m-1}}{(z_{14}\bar{z}_{14})} \frac{z_{12}^3}{z_{23}z_{34}z_{41}} \delta(r-\bar{r})\theta\left(r-1\right)   \nonumber \\  
     &\phantom{=} \times \prod_{i=1}^{4} \int_{0}^{\infty} \dd{\omega_{{i}}} \omega_{1}^{\Delta_1 + n } \omega_{2}^{\Delta_2 + n + m } \omega_{3}^{\Delta_3 + m - 2}\omega_{4}^{\Delta_{4}-3} \prod_{j=1}^{3} \delta\left(\omega_j - \chi_{j} \omega_{4}\right) \nonumber\\ 
     &= 4\, g_{10}^2 (\alpha')^{n+m} \frac{(z_{12}\bar{z}_{12})^{n}(z_{23}\bar{z}_{23})^{m-1}}{(z_{14}\bar{z}_{14})} \frac{z_{12}^3}{z_{23}z_{34}z_{41}} \delta(r-\bar{r}) \theta\left(r-1\right)  \nonumber \\ 
     &\phantom{=} \times \chi_{1}^{\Delta_1+n}\chi_{2}^{\Delta_2+n+m}\chi_{3}^{\Delta_{3}+m-2}  \int_{0}^{\infty} \frac{\dd{\omega_{4}}}{\omega_{4}} \omega_{4}^{\Delta+2(n+m)-4}. 
\label{eq:CelestialStringTreeExpansionPrelim}
\end{align}
Before continuing, we want to put special emphasis on the remaining integral in the last line which is known as a generalised Dirac delta distribution 
\begin{equation}
    \bdelta\left(i(\Delta-z)\right) \equiv \frac{1}{2\pi} \int_{0}^{\infty} \frac{\dd{\omega}}{\omega} \omega^{\Delta-z},
    \label{eq:DiracDeltaCelestial}
\end{equation}
in the sense that it allows for complex arguments. In the context of celestial holography, it was introduced \cite{donnayAsymptoticSymmetriesCelestial2020} as a means to allow for arbitrary analytically continued conformal scaling dimension $\Delta \in \mathbb{C} $ of conformal primaries and express them as superpositions of conformal primaries on the principle continuous series $\Delta = 1 + i\lambda, \, \lambda\in \mathbb{R} $. In this work we will naturally be inclined to consider analytically continued conformal scaling parameters $\beta = (4-\Delta)/2$ as well. And while (\ref{eq:DiracDeltaCelestial}) is a divergent object, further rigorous mathematical investigations \cite{Pano:2024eek} have shown that the divergence is distributional and tempered in the sense that they possess a well-defined finite action on functions inhabiting the Mellin transformed version of Schwartz space $\mathcal{S}(\mathbb{R}^+)$, the space of rapidly decreasing test functions. Moreover, it has previously been employed in works of other research fields \cite{brewsterGeneralizedDeltaFunctions2018}, wherein it has been demonstrated using contour integral techniques that it satisfies the usual sifting property
\begin{align}
    \int_{-\infty}^{\infty}  \dd{x} f(x) \bdelta(x-z_0) = f(z_{0}), \quad x\in \mathbb{R},
    \label{eq:SiftingPropertyDiracDelta} 
\end{align}
for points in the complex plane $z_{0} \in \mathbb{C}$.\footnote{The formal definition for the Dirac delta function given in \cite{brewsterGeneralizedDeltaFunctions2018} is 
\begin{align*}
    \bdelta(iz) = \int_{-\infty}^{\infty} \dd{p} e^{zp},
\end{align*}
which can easily be shown to be identical to (\ref{eq:DiracDeltaCelestial}) via change of variables $e^p = \omega$.
}
This is akin to performing a closed contour integral around a simple pole, the position of which is indicated by the argument of the delta distribution.\footnote{A possible method of removing even the distributional divergence entirely is introducing a UV-cutoff parameter $\Lambda$, resulting in integrals of the form $\int_{0}^{\Lambda} \dd{\omega} w^{\Delta-z-1} = \Lambda^{\Delta-z}/\left(\Delta-z\right)$. Note the appearance of the simple pole at $\Delta=z$, consistent with the previous statement.} We make constant implicit use of this property throughout this paper to evaluate expressions on the support of this delta distribution. For the case at hand we have
\begin{align}
    \int_{0}^{\infty} \frac{\dd{\omega_{4}}}{\omega_{4}} \omega_{4}^{\Delta+2(n+m)-4} = \frac{1}{2} \int_{0}^{\infty} \dd{x} x^{(n+m-\beta)-1} = \pi \, \bdelta\left(i(n+m-\beta)\right).
\end{align}
This imposes $n+m=\beta$, allowing us to simplify our final expression to 
\begin{align}
    \mathscr{J}_{\text{tree}}(n,m) &= 4 \, g_{10}^2 (\alpha')^{\beta} K\left(h_i,\bar{h}_i\right) r^{\frac{5-\beta}{3}}\left(r-1\right)^{\frac{2-\beta}{3}}  \delta(r-\bar{r}) 
        \theta\left(r-1\right) I_{\text{tree}}(n,m), 
        \label{eq:CelestialStringTreeExpansionIntegral}
\end{align}
where we dropped an overall phase $(-1)^{\frac{4}{3}(1+\beta)+i(\lambda_2+\lambda_3) }$ for simplicity again and, in analogy with $I\left(r,\beta\right)$ in  (\ref{eq:TreeEnergyIntegral}), we defined the remaining energy integral as 
\begin{align}
	I_{\text{tree}}(n,m) = \pi \, r^{\beta-m} \bdelta(i(n+m-\beta)).
	\label{eq:CelestialTreeExpansionEnergyIntegral}
\end{align}
As a first sanity check, note that evaluating the expression on the support of the generalised delta distribution precisely reproduces the same overall $(\alpha')^{\beta}$ factor expected from celestial string tree amplitudes. This holds true at \emph{any} order in the low-energy expansion since the $\alpha'$ factor in each term generally reads $(\alpha')^{n+m}$ as seen in equation (\ref{eq:CelestialStringTreeExpansionPrelim}). Furthermore, note that all terms in (\ref{eq:CelestialStringTreeExpansionIntegral}) in front of the energy integral $I_{\text{tree}}(n,m) $ match the ones in the fully fledged result (\ref{eq:CelestialStringTree}). The remaining task at hand is performing a term by term comparison of the energy integrals $I\left(r,\beta\right)$ and $I_{\text{tree}}(n,m)$. To this end we need to take a more detailed look at the $\alpha'$-expansion at tree level given in  (\ref{eq:ExpansionVeneziano}). Since we defined $n$ and $m$ to simply count the powers of the appearing Mandelstam variables $s_{12}$ and $s_{23}$ respectively, we see that the zeroth order term in the expansion simply corresponds to the case $(n,m) = (0,0)$ and therefore $I_{\text{tree}}(0,0) = \pi\bdelta(i\beta)$.\footnote{Here we used the property $\bdelta(ia\beta) = \abs{a}^{-1}\bdelta(i\beta) $ for $a = -1 $ which can be derived by using the integral representation (\ref{eq:DiracDeltaCelestial}) and performing a change of variables $\omega = x^{a}$.} Using the support of the delta distribution once again, we obtain 
\begin{align}
	\mathscr{J}_{\text{tree}}(0,0) &= 4 \pi \, g_{10}^2 \, \bdelta(i\beta) K\left(h_i,\bar{h}_i\right) r^{\frac{5}{3}}\left(r-1\right)^{\frac{2}{3}}\delta(r-\bar{r})\theta\left(r-1\right),
\end{align}
faithfully reproducing the YM contribution to the full celestial string tree level amplitude. As for the higher order results, note that as previously mentioned there is no linear term in $\alpha'$ in the expansion and correspondingly no terms with $(n,m)=\{(1,0),(0,1)\}$. Indeed, the infinite stringy tower of states will be indexed starting with $(n,m)=(1,1)$ and terms for a given fixed number $n+m$ will be grouped together accompanied by a product $\zeta_{n+m} \, \bdelta(i\left(n + m - \beta\right)) = \zeta_{\beta} \, \bdelta(i\left(n + m - \beta\right)) $ as dictated by (\ref{eq:ExpansionVeneziano}) in conjunction with (\ref{eq:CelestialTreeExpansionEnergyIntegral}). Concretely, the celestial stringy correction terms to the YM amplitude read
\begin{multline}
\mathscr{C}_{\text{tree}}\left(r,\beta \right) 
=	  \zeta_{\beta} \, r^{\beta-1} \, \Bigl\{
           \bdelta(i\left(\beta -2 \right))  + \left[1-r^{-1}\right] \, \bdelta(i\left(\beta -3 \right)) 
           \\ 
    +  \left[1 -\left(4r\right)^{-1} + r^{-2}\right] \, \bdelta(i\left(\beta -4 \right)) + \cdots
        \Bigr\}.
    \label{eq:StringyCorrectionTermsTreeLevel} 	
\end{multline}
Note that the sifting property of the generalised delta distribution allowed all zeta functions $\zeta_{2},\zeta_{3},\ldots$ to be factored out as a pre-factor $\zeta_{\beta}$ similar to how the pre-factor $\left(\alpha'\right)^{\beta}$ for the entire tree level amplitude manifested. The delta distributions mark the location of the poles of the full amplitude in the complex $\beta$-plane at $\beta=n, \, \forall \, n \in \mathbb{N}_{\geq 2}$ and their accompanying terms represent their corresponding residues as suggested on general grounds in \cite{arkani-hamedCelestialAmplitudesUV2021}.\footnote{Note that in \cite{arkani-hamedCelestialAmplitudesUV2021}, the authors used the convention $\beta= \sum_{k=1} \left(\Delta_{k}-1\right) = \Delta -4 $ which differs from ours $\Delta = 4 - 2\beta$ by a factor of $-1/2$.} A noteworthy property of the string correction terms $\mathscr{C}_{\text{tree}}\left(r,\beta \right)$ is the symmetry $ r^{\beta-m} =  r^{n}$ implied by the delta distributions for all terms in the expansion, allowing for an alternative expression
\begin{multline}
\mathscr{C}_{\text{tree}}\left(r,\beta \right) 
=	  \zeta_{\beta} \, r \, \Bigl\{
           \bdelta(i\left(\beta -2 \right))  + \left[1-r\right] \, \bdelta(i\left(\beta -3 \right)) 
           \\ 
    +  \left[1 - \left(1/4\right) r + r^{2}\right] \, \bdelta(i\left(\beta -4 \right))
       +\cdots \Bigr\}.
    \label{eq:StringyCorrectionTermsTreeLevelAlternative} 	
\end{multline}
However, the stringy correction terms with increasingly negative powers of $r$ as seen in (\ref{eq:StringyCorrectionTermsTreeLevel}) represent the more natural choice, given that $r=\sin(\theta/2)^{-2}>1$ for physical scattering angles $\theta \in [0,2\pi] \in \mathbb{R}$ in the center of mass frame, allowing for a recovery of YM results in the kinematic limit of large $r$ where the stringy correction terms are negligible compared to the field theory term. By contrast, having increasingly positive powers of $r$ as suggested by (\ref{eq:StringyCorrectionTermsTreeLevelAlternative}) would require taking the limit of $r\to 0$ to recover the YM contribution, thereby violating the s-channel constraint $r>1$. In conclusion, the full amplitude expression obtained by this ansatz reads
\begin{multline}
\tilde{\mathscr{A}}_{\text{String}}^{\, \text{tree}}(1^-,2^-,3^+,4^+) = 4 \pi \, g_{10}^2 (\alpha')^{\beta} K\left(h_i,\bar{h}_i\right) r^{\frac{5-\beta}{3}}\left(r-1\right)^{\frac{2-\beta}{3}}  \delta(r-\bar{r})\theta\left(r-1\right) \\ 
\times\left( 
\bdelta(i\beta)  + \mathscr{C}_{\text{tree}}\left(r,\beta \right) 
\right).
\label{eq:TreeLevelWholeResult}   	
\end{multline}
\tikzset{every picture/.style={line width=0.75pt}}
\begin{figure}[t]
    \centering
    \begin{tikzpicture}[x=0.75pt,y=0.75pt,yscale=-1,xscale=1,,baseline=(current bounding box.center)]
\draw [color={rgb, 255:red, 173; green, 173; blue, 173 }  ,draw opacity=0.7 ][line width=1.5]  [dash pattern={on 5.63pt off 4.5pt}]  (279.6,279.94) -- (385.75,279.94) ;
\draw [color={rgb, 255:red, 173; green, 173; blue, 173 }  ,draw opacity=0.7 ][line width=1.5]    (210,280) -- (279.6,279.94) ;
\draw [color={rgb, 255:red, 173; green, 173; blue, 173 }  ,draw opacity=0.7 ][line width=1.5]    (140.4,280.06) -- (210,280) ;
\draw  [color={rgb, 255:red, 255; green, 255; blue, 255 }  ,draw opacity=0 ][fill={rgb, 255:red, 208; green, 2; blue, 27 }  ,fill opacity=1 ] (135.23,280.06) .. controls (135.23,277.21) and (137.55,274.9) .. (140.4,274.9) .. controls (143.25,274.9) and (145.57,277.21) .. (145.57,280.06) .. controls (145.57,282.92) and (143.25,285.23) .. (140.4,285.23) .. controls (137.55,285.23) and (135.23,282.92) .. (135.23,280.06) -- cycle ;
\draw  [draw opacity=0] (151.68,288.26) .. controls (149.14,291.74) and (145.04,294) .. (140.4,294) .. controls (132.7,294) and (126.46,287.76) .. (126.46,280.06) .. controls (126.46,272.37) and (132.7,266.13) .. (140.4,266.13) .. controls (145.38,266.13) and (149.76,268.74) .. (152.22,272.68) -- (140.4,280.06) -- cycle ; \draw   (151.68,288.26) .. controls (149.14,291.74) and (145.04,294) .. (140.4,294) .. controls (132.7,294) and (126.46,287.76) .. (126.46,280.06) .. controls (126.46,272.37) and (132.7,266.13) .. (140.4,266.13) .. controls (145.38,266.13) and (149.76,268.74) .. (152.22,272.68) ;  
\draw   (123.08,284.8) -- (126.36,278.25) -- (129.64,284.8) ;
\draw  [color={rgb, 255:red, 255; green, 255; blue, 255 }  ,draw opacity=0 ][fill={rgb, 255:red, 208; green, 2; blue, 27 }  ,fill opacity=1 ] (204.83,280) .. controls (204.83,277.15) and (207.15,274.83) .. (210,274.83) .. controls (212.85,274.83) and (215.17,277.15) .. (215.17,280) .. controls (215.17,282.85) and (212.85,285.17) .. (210,285.17) .. controls (207.15,285.17) and (204.83,282.85) .. (204.83,280) -- cycle ;
\draw  [color={rgb, 255:red, 255; green, 255; blue, 255 }  ,draw opacity=0 ][fill={rgb, 255:red, 208; green, 2; blue, 27 }  ,fill opacity=1 ] (274.43,279.94) .. controls (274.43,277.08) and (276.75,274.77) .. (279.6,274.77) .. controls (282.45,274.77) and (284.77,277.08) .. (284.77,279.94) .. controls (284.77,282.79) and (282.45,285.1) .. (279.6,285.1) .. controls (276.75,285.1) and (274.43,282.79) .. (274.43,279.94) -- cycle ;
\draw  [draw opacity=0] (198.05,272.82) .. controls (200.49,268.77) and (204.93,266.06) .. (210,266.06) .. controls (215.16,266.06) and (219.66,268.86) .. (222.07,273.03) -- (210,280) -- cycle ; \draw   (198.05,272.82) .. controls (200.49,268.77) and (204.93,266.06) .. (210,266.06) .. controls (215.16,266.06) and (219.66,268.86) .. (222.07,273.03) ;  
\draw    (152.09,272.47) -- (198.25,272.47) ;
\draw   (172.26,269.11) -- (178.82,272.39) -- (172.26,275.67) ;
\draw  [draw opacity=0] (221.42,287.99) .. controls (218.9,291.59) and (214.72,293.94) .. (210,293.94) .. controls (205.28,293.94) and (201.1,291.59) .. (198.58,287.99) -- (210,280) -- cycle ; \draw   (221.42,287.99) .. controls (218.9,291.59) and (214.72,293.94) .. (210,293.94) .. controls (205.28,293.94) and (201.1,291.59) .. (198.58,287.99) ;  
\draw    (151.68,288.26) -- (198.72,288.26) ;
\draw  [draw opacity=0] (267.53,272.97) .. controls (269.94,268.8) and (274.44,266) .. (279.6,266) .. controls (284.76,266) and (289.26,268.8) .. (291.67,272.97) -- (279.6,279.94) -- cycle ; \draw   (267.53,272.97) .. controls (269.94,268.8) and (274.44,266) .. (279.6,266) .. controls (284.76,266) and (289.26,268.8) .. (291.67,272.97) ;  
\draw  [draw opacity=0] (291.02,287.93) .. controls (288.5,291.53) and (284.32,293.88) .. (279.6,293.88) .. controls (274.88,293.88) and (270.7,291.53) .. (268.18,287.93) -- (279.6,279.94) -- cycle ; \draw   (291.02,287.93) .. controls (288.5,291.53) and (284.32,293.88) .. (279.6,293.88) .. controls (274.88,293.88) and (270.7,291.53) .. (268.18,287.93) ;  
\draw    (221.42,287.99) -- (268.18,287.99) ;
\draw    (291.67,272.97) -- (350,272.97) ;
\draw    (221.59,272.72) -- (267.75,272.72) ;
\draw   (241.76,269.36) -- (248.32,272.64) -- (241.76,275.92) ;

\draw   (177.82,291.42) -- (171.26,288.14) -- (177.82,284.86) ;
\draw   (248.32,291.17) -- (241.76,287.89) -- (248.32,284.61) ;
\draw   (206.26,263.11) -- (212.82,266.39) -- (206.26,269.67) ;
\draw    (291.02,287.93) -- (350.25,287.93) ;
\draw  [dash pattern={on 0.84pt off 2.51pt}]  (350,272.97) -- (371,272.97) ;
\draw  [dash pattern={on 0.84pt off 2.51pt}]  (350.25,287.93) -- (371.25,287.93) ;
\draw   (213.57,297.17) -- (207.01,293.89) -- (213.57,290.61) ;
\draw   (283.57,296.92) -- (277.01,293.64) -- (283.57,290.36) ;
\draw   (325.57,291.17) -- (319.01,287.89) -- (325.57,284.61) ;
\draw   (276.51,262.86) -- (283.07,266.14) -- (276.51,269.42) ;
\draw   (318.76,269.61) -- (325.32,272.89) -- (318.76,276.17) ;
\draw (142.4,283.46) node [anchor=north west][inner sep=0.75pt]  [font=\tiny]  {$2$};
\draw (212,283.4) node [anchor=north west][inner sep=0.75pt]  [font=\tiny]  {$3$};
\draw (281.6,283.34) node [anchor=north west][inner sep=0.75pt]  [font=\tiny]  {$4$};
\draw (113.67,266.73) node [anchor=north west][inner sep=0.75pt]  [font=\footnotesize]  {$\mathcal{C}$};
\end{tikzpicture}
    \caption[]{Contour $\mathcal{C}$ enclosing the poles \tikz[x=0.75pt,y=0.75pt,yscale=-1.00,xscale=1.00,baseline={([yshift=-3pt]current bounding box.center)}]{\draw  [color={rgb, 255:red, 255; green, 255; blue, 255 },draw opacity=0][fill={rgb, 255:red, 208; green, 2; blue, 27 }  ,fill opacity=1 ] (59.48,95.08) .. controls (59.48,92.23) and (61.8,89.92) .. (64.65,89.92) .. controls (67.5,89.92) and (69.82,92.23) .. (69.82,95.08) .. controls (69.82,97.94) and (67.5,100.25) .. (64.65,100.25) .. controls (61.8,100.25) and (59.48,97.94) .. (59.48,95.08) -- cycle ;}  of $\pi\sin(\pi\beta)^{-1}$ at $n\in\mathbb{N}_{\geq 2}$ in complex $\beta$-plane.}
\label{fig:SinePoleStructure} 
\end{figure}
We can now proceed to display the connection between the string correction terms of the complete result in (\ref{eq:TreeEnergyIntegral2}) and those obtained via $\alpha'$-expansion (\ref{eq:StringyCorrectionTermsTreeLevel}) highlighting the pole structure in the complex $\beta$-plane. An obvious source for poles in the full result is the $\sin(\pi\beta)^{-1}$ factor yielding simple poles $\forall \, \beta = n \in \mathbb{Z}$. A nontrivial potential source for poles is the MZV $\zeta\left(\beta-k,\{1\}^{k}\right)$. It is defined as an absolutely convergent and hence pole free series in the domain $ \real{\left(\beta\right)}-k > 1 $ which, in conjunction with $k \in \mathbb{N}_{\geq 0}$, is equivalent to $\real{\left(\beta\right)} > 1$. In this domain, the only poles of the amplitude originate from the sine function. They are positioned at $\beta = n \in \mathbb{N}_{\geq 2}$ in exact agreement with the pole structure emerging in (\ref{eq:StringyCorrectionTermsTreeLevel}). In order to pick up these contributions, we used the inverse Mellin contour $\mathcal{C}$ in the complex $\beta$ plane as used in \cite{donnayAsymptoticSymmetriesCelestial2020} consisting of an infinite line along the imaginary axis $(c-i\infty,c+i\infty)$. By construction, the real component $c$ has to be defined on the common strip of holomorphy $S_{c} \supset \{\beta \in \mathbb{C} \,| \, 1<\real{\left(\beta\right)}<2$\}  \cite{Pano:2024eek}, which coincides precisely with the domain where the MZV is absolutely convergent $\real{\left(\beta\right)} >1$ but where the contour is still to the left of the first pole in that domain at $\beta=2$. We continue by closing the contour the right as depicted in Figure \ref{fig:SinePoleStructure}. Since the  rest of the amplitude is analytic and we deal exclusively with simple poles, the other terms will factor out evaluated on the positions of the poles by the residue theorem. Let us therefore focus on the relevant part of the contour integral for simplicity an compute
\begin{align}
	\oint_{\mathcal{C}} \dd{\left(-i\beta\right)} \frac{\pi}{\sin(\pi\beta)} = 2\pi \sum_{n=2}^{\infty} \Res\left[ \, \frac{\pi}{\sin(\pi\beta)},\beta=n\right] =  2\pi \sum_{n=2}^{\infty} (-1)^{n+1}.
\end{align}
Note that we obtained an extra sign factor because the contraction of our contour of choice $\mathcal{C}$ leads to clockwise, i.e. negatively oriented circles around the poles. Plugging in this result into the stringy correction terms of (\ref{eq:TreeEnergyIntegral2}), we obtain the intermediate expression
\begin{align}
	\pi  \, r^{\beta-1} \sum_{n=2}^{\infty}   \sum_{k=0}^{\infty} \left(-r\right)^{-k}\zeta\left(n-k,\{1\}^{k}\right),
	\label{eq:IntermediateExpansionTree}
\end{align}
which, for any fixed $n$ should reproduce the terms in our expansion at the pole $\beta=n$ as indicated by $\bdelta(i\left(\beta -n \right))$. For this purpose, we note some known relations \cite{gilMultipleZetaValues2017} between different MZVs 
\begin{align}
	\zeta_3 = \zeta\left(2,1\right), \qquad 
	\zeta_4 = 4 \, \zeta\left(3,1\right) = \zeta\left(2,1,1\right).
	\label{eq:MZVIdentitiesExplicit}  
\end{align}
Considering the first three terms $n=\{2,3,4\}$, performing the sum over $k$ by truncating terms violating the constraint for absolute convergence $ \real{\left(\beta\right)}-k > 1 $, we obtain\footnote{In this notation, $\{1\}^{k}$ for $k=0$ is to be understood as the empty set $ \{1\}^{0} = \varnothing$ and therefore reduces the MZV to the regular zeta value with a single argument.}
\begin{align}
	 \sum_{k=0}^{\infty} \left(-r\right)^{-k}\zeta\left(2-k,\{1\}^{k}\right) &= \zeta_{2}, \nonumber \\
	 \sum_{k=0}^{\infty} \left(-r\right)^{-k}\zeta\left(3-k,\{1\}^{k}\right) &= \zeta_{3} + \left(-r\right)^{-1}\zeta\left(2,1\right) = \zeta_{3} \left(1-r^{-1}\right), \label{eq:MZVSumEvals} \\ 
	  \sum_{k=0}^{\infty} \left(-r\right)^{-k}\zeta\left(4-k,\{1\}^{k}\right) &= \zeta_{4} + \left(-r\right)^{-1}\zeta\left(3,1\right) +  \left(-r\right)^{-2}\zeta\left(2,1,1\right) = \zeta_{4} \left(1-(4r)^{-1} + r^{-2}\right), \nonumber
\end{align} 
where we employed (\ref{eq:MZVIdentitiesExplicit}) in the last step of the last two lines. Combining the results in (\ref{eq:MZVSumEvals}) with (\ref{eq:IntermediateExpansionTree}) reproduces the results of the $\alpha'$-expansion ansatz (\ref{eq:StringyCorrectionTermsTreeLevel}).
 
To consider the as of yet unaccounted poles and their residues of the $\pi\sin(\pi \beta)^{-1}$ factor at $\beta \in \mathbb{Z}_{\leq 1}$, one needs to analytically continue the MZV to a meromorphic function on the entire complex plane \cite{zhaoAnalyticContinuationMultiple2000} and analyse the resulting overall expression. For the particular MZV of interest, it reads  
\begin{align}
	\zeta\left(\beta-k,\{1\}^{k}\right) = \frac{[\Gamma\left(\beta-k-1\right)]^{k}}{\left(\beta-k-1\right)} \xi\left(\beta-k,\{1\}^{k}\right),
	\label{eq:ExplicitAnalyticContinuationMZV}
\end{align}
where $\xi$ is an \emph{entire} function on $\mathbb{C}$.\footnote{In general, the MZV is analytically continued to the entire complex hyperplane  $\mathbb{C}^{k+1}$, but in this special case all but one of the $k+1$ complex variables are evaluated at the fixed value $1$, reducing the space down to $\mathbb{C}$.} The additional sources of potential poles include the vanishing denominator at $\beta=k+1$ and the $k$ identical Gamma functions, each yielding simple poles when their argument evaluates to a non-positive integer, i.e. $\beta = k+1-\ell$ for the stringy tower sum index $k\in\mathbb{N}_{k\geq 0}$ and a generic non-negative integer $\ell \in \mathbb{N}_{\geq 0}$. It follows that for any given string correction term in the infinite series (\ref{eq:TreeEnergyIntegral2}) with a fixed value of $k$, the analytically continued MZV by itself will provide an infinite set of poles starting at $\beta = k+1$ and then moving towards the left along the real axis in integer steps, inevitably entering the previously unexplored domain $\real{\left(\beta\right)} \leq  1$, including the negative real line. This is meaningful because, as pointed out in \cite{arkani-hamedCelestialAmplitudesUV2021}, IR and UV contributions to a celestial amplitudes of UV-complete theories can be distinguished by checking if their poles are located either on the positive or negative real axis in the complex $\beta$-plane. This suggests that studying the residues of the poles located on the negative real $\beta$ axis using the analytic continuation might lead to new insights in the UV behaviour of the amplitude. We want to emphasize that these residues do not seem to vanish, at least not identically, as the simple zeros of the MZV are not located at these values (\ref{eq:ExplicitAnalyticContinuationMZV}). Indeed, determining the full set of trivial zeros and deriving a possible functional equation is still an open problem. We leave this task for future research endeavours.  

Another connection between the string formfactor and its $\alpha'$-expansion has been pointed out very recently in the context of Carrollian strings \cite{stiebergerCarrollianAmplitudesStrings2024} using results found in \cite{borweinEvaluationsKfoldEuler1996}. One can express the string formfactor as 
\begin{align}
	F_{I}(\alpha' s,\alpha' u) = 1 + \sum_{k,\ell\geq0} c_{k\ell} \left(-1\right)^{\ell}\left(\alpha'\right)^{k+\ell+2}  \left(s_{12}\right)^{k+1} \left(s_{23}\right)^{\ell+1},
	\label{eq:DrinfeldExpansionAnsatz}
\end{align}
where the expansion coefficients $c_{k\ell}$ are MZVs satisfying a special case of the Ohno duality relation
\begin{align}
	c_{k\ell} = c_{\ell k} = \zeta\left(\ell + 2,\{1\}^{k}\right) = \zeta\left(k + 2,\{1\}^{\ell}\right), \qquad \forall \, \ell,k \in \mathbb{N}_{\geq 0}.
\end{align}
Note that relation implies  $\zeta_3 = \zeta\left(2,1\right)$ and $\zeta_4 = \zeta\left(2,1,1\right)$ as previously mentioned in (\ref{eq:MZVIdentitiesExplicit}). From the point of view of the Mellin transform, the term in (\ref{eq:DrinfeldExpansionAnsatz}) leads to a virtually identical computation as the one in (\ref{eq:CelestialStringTreeExpansionPrelim}) up to some signs and an index shift $n \to k +1$ and $m \to \ell +1$. Hence, we will not repeat this computation. Suffices it to say that one obtains the same result as displayed in (\ref{eq:TreeLevelWholeResult}) in the end.

Another noteworthy property of the generalised delta distribution that has been pointed out in both papers \cite{donnayAsymptoticSymmetriesCelestial2020,brewsterGeneralizedDeltaFunctions2018} is that when the imaginary part of its argument goes to zero, it reduces to the regular delta distribution. Since by construction $n,m \in \mathbb{N}_{\geq 0} $, the imaginary part is simply $\imaginary(\beta) = -\lambda/2$. Taking the limit $\beta \to 0$ is therefore akin to taking $\lambda \to 0$, which physically speaking corresponds precisely to the \emph{conformal soft limit} \cite{fanSoftLimitsYangMills2019}.

\subsection{One Loop Planar Annulus and M{\"o}bius Amplitude}
\label{sec:PlanarLoopLevelCelestialExpansion}

For the planar loop sector, the computation needed to celestialize the $\alpha'$-expansion is very similar to that at tree level. The expansions of the cylinder and M{\"o}bius amplitude can be related by flipping the sign of the modular parameter $q \to -q$, which is irrelevant at the level of the Mellin transform. We will therefore first focus on the cylinder contribution. The object of interest is
\begin{align}
	\tilde{\mathcal{A}}_{\text{Cyl}}\left(1,2,3,4\right) = \frac{g_{10}^{2}}{\alpha'} \prod_{i=1}^{4} \int_{0}^{\infty} \dd{\omega_{{i}}}^{\Delta_i -1} s_{12}s_{23}\mathcal{A}_{\text{YM}}^{\text{tree}}(1,2,3,4) \int_{0}^{1} \frac{\dd{q}}{q} I_{1234}(q).
\end{align}
The $\alpha'$-expansion (\ref{eq:PlanarPunctureIntegral}) is again given in terms of powers of Mandelstam variables like at tree level but with accompanying $q$-dependent eMZVs instead of constant MZVs. These will need to be accounted for later. However, for the Mellin transform they are just bystanders again as they are independent of all energies $\omega_{i}$. Inspired by this, we exchange the order of integration leading to a definition of the energy integrals of the form
\begin{align}
	\mathscr{J}_{\text{Loop}}\left(n,m\right) = \frac{g_{10}^{2}}{(\alpha')^{3}} \prod_{i=1}^{4} \int_{0}^{\infty} \dd{\omega_{{i}}} \omega_{i}^{\Delta_i -1} \mathcal{A}_{\text{YM}}^{\text{tree}}(1^-,2^-,3^+,4^+)(\alpha' s_{12})^{n+1}(\alpha' s_{23})^{m+1}. 
\end{align}
This is of course strikingly similar to the energy integral at tree level (\ref{eq:CelestialStringTreeExpansionPrelim}). Indeed, by direct comparison one can immediately read off
\begin{align}
	\mathscr{J}_{\text{Loop}}(n,m) = \left(\alpha'\right)^{-3} \mathscr{J}_{\text{Tree}}(n+1,m+1), 
\end{align}
or more explicitly  
\begin{align}
    \mathscr{J}_{\text{Loop}}(n,m) &= 4 \, g_{10}^2 (\alpha')^{\beta-3} K\left(h_i,\bar{h}_i\right) r^{\frac{5-\beta}{3}}\left(r-1\right)^{\frac{2-\beta}{3}}  \delta(r-\bar{r}) 
        \theta\left(r-1\right) I_{\, \text{Loop}}(n,m), 
\end{align}
with the expansion term-dependent integral reading
\begin{align}
	I_{\, \text{Loop}}(n,m) = I_{\text{tree}}(n+1,m+1) = \pi \, r^{\beta-m-1} \bdelta(i(2+n+m-\beta)).
	\label{eq:PlanarLoopEnergyIntegral}
\end{align}
Using the explicit amplitude expansion amplitude expansion (\ref{eq:PlanarPunctureIntegral}), we ultimately obtain
\begin{align}
	\tilde{\mathcal{A}}_{\text{Cyl}}\left(1,2,3,4\right) &= 4 \, g_{10}^2 (\alpha')^{\beta-3} K\left(h_i,\bar{h}_i\right) r^{\frac{5-\beta}{3}}\left(r-1\right)^{\frac{2-\beta}{3}}  \delta(r-\bar{r}) 
        \theta\left(r-1\right) \nonumber \\ 
        &\phantom{=} \times \pi \, r^{\beta-1} \int_{0}^{1} \frac{\dd{q}}{q} \bigg\{ 
        \varphi(0,0,0)\, \bdelta(i(\beta-2)) -  \, 2\, \varphi(0,1,0,0) \, \left[1+ r^{-1}\right] \bdelta(i(\beta-3)) \nonumber \\
        &\phantom{=} + \left[ \, 2 \, \varphi(0,1,1,0,0)\,  \left( 1  + r^{-2} \right)    
 - \, 2 \,  \varphi(0,1,0,1,0) \, r^{-1} \,\right] \bdelta(i(\beta-4)) \nonumber \\ 
    	&\phantom{=}
    	+ \left[ \sigma_5 \, (1+2 r^{-1} + 2 r^{-2} +r^{-3})
\, + \, \sigma_{2,3} \, r^{-1}(1+r^{-1})\right] \bdelta(i(\beta-5))  
        + \cdots \bigg\}. 
        \label{eq:ExpansionPlanarCylinder}
\end{align}
Let us analyse the properties of this intermediate result. The $\alpha'$-dependence has again been reduced to that of a simple global factor $(\alpha')^{\beta-3}$ consistent with the results in section \ref{sec:CelestialLoopIntegrands}. The poles of the cylinder loop correction in the $\beta$-plane are located at $\beta = n \in \mathbb{N}_{\geq 2} $, precisely coinciding with the positions of the stringy correction terms to the YM amplitude at tree level (\ref{eq:StringyCorrectionTermsTreeLevel}). Moreover,  it also introduces the same additional inverse powers of the conformal cross ratios of at least $r^{\beta-1}$, allowing for a consistent reproduction of the field theory amplitude in the aforementioned  limit $r \to \infty$.  

The last remaining challenge is the remaining $\int_{0}^{1} \dd{q}/q$ integrals with the exponentiated modular parameter $q=e^{-2\pi t}$. The only $q$-dependent terms in our expression are the eMZVs. They generally can be expressed as some constants plus a double sum of the form  
\begin{align}
	\Phi\left(a,b;q\right) = \sum_{m,n\geq1} \, \frac{q^{mn}}{m^a n^b}.
	\label{eq:eMZVExpansion}   
\end{align}
All explicit eMZV examples relevant for this paper have been gathered in Appendix \ref{sec:eMZVs}. In the planar case, the moduli space integrals necessitate a regularisation method, lest we end up a divergent harmonic series from (\ref{eq:eMZVExpansion}) if either $a=0$ or $b=0$. We employ the $\varepsilon$-regularisation seen for instance in \cite{hoheneggerMonodromyRelationsHigherLoop2017}, which amounts to adding a factor $t^{\varepsilon}$ with a parameter $\epsilon > -1$ in the integrand. Performing a change of variables from $q$ to $t$, one can compute 
\begin{align}
	 \int_{0}^{\infty} \dd{t} t^{\varepsilon} \, q^{mn} =  \int_{0}^{\infty} \dd{t} t^{\varepsilon} \, e^{-2\pi mnt} = \frac{1}{(2\pi mn)^{1+\varepsilon}} \int_{0}^{\infty} \dd{u} u^{\varepsilon} e^{-u} =   \frac{\Gamma\left(1+\varepsilon\right)}{(2\pi mn)^{1+\varepsilon}},  
\end{align}
evaluating the remaining sums, we have
\begin{multline}
	 \int_{0}^{\infty} \dd{t} t^{\varepsilon} \, \Phi\left(a,b;q\right)
    = \frac{\Gamma(1+\varepsilon)}{(2\pi)^{1+\varepsilon}}  \sum_{m=1}^{\infty} \frac{1}{m^{1+a+\varepsilon}} \sum_{n=1}^{\infty} \frac{1}{n^{1+b+\varepsilon}}
    = \frac{\Gamma(1+\varepsilon)}{(2\pi)^{1+\varepsilon}}   \zeta_{1+b+\varepsilon} \zeta_{1+a+\varepsilon}.
    \label{eq:FirstRegularizationIntegral}
\end{multline}
For the M{\"o}bius strip, performing the sums requires slightly more effort due to the sign flip $q \to -q$. We compute 
\begin{align}
    \int_{0}^{\infty} \dd{t} t^{\varepsilon} \, \Phi\left(a,b;-q\right) &= \frac{ \Gamma\left(1+\varepsilon\right)}{(2\pi)^{1+\varepsilon}}   \sum_{n= 1}^{\infty} \frac{1}{n^{1+b+\varepsilon}} \sum_{m = 1}^{\infty} \frac{(-1)^{nm}}{m^{1+a+\varepsilon}} \nonumber \\  
    &= \frac{ \Gamma\left(1+\varepsilon\right)}{(2\pi)^{1+\varepsilon}}   \sum_{n=1}^{\infty} \frac{\mathcal{L}i_{1+a+\varepsilon}\left((-1)^{n})\right)}{n^{1+b+\varepsilon}}   \nonumber \\ 
    &= \frac{ \Gamma\left(1+\varepsilon\right)}{(2\pi)^{1+\varepsilon}} \left\{ 
     \sum_{n=1}^{\infty} \frac{\mathcal{L}i_{1+a+\varepsilon}\left(-1\right)}{(2n-1)^{1+b+\varepsilon}} + \sum_{n=1}^{\infty} \frac{\mathcal{L}i_{1+a+\varepsilon}\left(1\right)}{(2n)^{1+b+\varepsilon}}
    \right\} \nonumber \\ 
    &=  \frac{ \Gamma\left(1+\varepsilon\right)}{(2\pi)^{1+\varepsilon}} \zeta_{1+a+\varepsilon} \, \zeta_{1+b+\varepsilon} \left\{ 
    \left(2^{-(a+\varepsilon)}-1\right) \left(1-2^{-(1+b+\varepsilon)}\right) + 2^{-(1+b+\varepsilon)}
    \right\} \nonumber \\ 
    &=  \left\{ 
    \left(2^{-(a+\varepsilon)}-1\right) \left(1-2^{-(1+b+\varepsilon)}\right) + 2^{-(1+b+\varepsilon)}
    \right\}  \int_{0}^{\infty} \dd{t} t^{\varepsilon} \, \Phi\left(a,b;q\right), 
\end{align}
where we used the basic definition of the polylogarithm 
\begin{align}
	\mathcal{L}_{s}\left(z\right) = \sum_{k=1}^\infty \frac{z^k}{k^s},
\end{align}
split up the sum in even and uneven integers, performed an index shift and used some straightforward identities
\begin{align}
	\mathcal{L}i_{s}\left( 1 \right) = \zeta_{s}, \qquad \mathcal{L}i_{s}\left( -1 \right) = \left(2^{1-s}-1\right)\zeta_s,
\end{align}
and noted that the end result agrees with the orientable case up to an overall numerical factor. Because of the symmetry $\Phi\left(a,b;q\right) = \Phi\left(b,a;q\right)$, one only needs to consider two instead of four cases when performing the $\varepsilon$-expansion. The relevant cases are for the cylinder read,
\begin{align}
	\int_{0}^{\infty} \dd{t} t^{\varepsilon} \, \Phi\left(a,b;q\right) &= \frac{1}{2\pi}\zeta_{1+a}\zeta_{1+b}  + \mathcal{O}\left(\varepsilon^{1}\right), \\
		 \int_{0}^{\infty} \dd{t} t^{\varepsilon} \, \Phi\left(a,0;q\right) &= \frac{1}{2\pi}\left\{ 
    \left(\frac{1}{\varepsilon} - \log(2\pi)\right) \zeta_{1+a} + \zeta'_{1+a}
    \right\} + \mathcal{O}\left(\varepsilon^{1}\right), 
\end{align}
with the derivative of the zeta value $\zeta'_{s} = - \sum_{s=1}^{\infty} \frac{\log n}{n^{s}}$. Expectedly, the expansion for the M{\"o}bius strip yields similar results
\begin{align}
	\int_{0}^{\infty} \dd{t} t^{\varepsilon} \, \Phi\left(a,b;-q\right) &= \frac{1}{2\pi}\zeta_{1+a}\zeta_{1+b} \left( 
    2^{-a}+2^{-b}-2^{-(1+a+b)}-1
    \right)  + \mathcal{O}\left(\varepsilon^{1}\right), \\
    \int_{0}^{\infty} \dd{t} t^{\varepsilon} \, \Phi\left(a,0;-q\right) &=  \frac{2^{-(a+1)}}{2\pi}\left\{
    \left(\frac{1}{\varepsilon} -\log (2\pi) - 2^{a+1} \log (2) 
    \right)\zeta_{1+a} +\zeta'_{1+a} 
    \right\} +  \mathcal{O}\left(\varepsilon^{1}\right).
\end{align}
This takes care of the q-dependent parts of the eMZVs. The divergent integrals over the constants in the eMZVs,  see for instance (\ref{eq:eMZVDivergentExample1}), will cancel with those of the M{\"o}bius strip as discussed before at the end of section \ref{sec:StringLoopAmplitudes}. Let us therefore focus solely on the finite pieces of the cylinder amplitude. The first few terms read 
\begin{align}
	\tilde{\mathcal{A}}_{\text{Cyl}}\left(1,2,3,4\right)|_{\text{fin}} &= 2 \, g_{10}^2 (\alpha')^{\beta-3} K\left(h_i,\bar{h}_i\right) r^{\frac{5-\beta}{3}}\left(r-1\right)^{\frac{2-\beta}{3}}  \delta(r-\bar{r}) 
        \theta\left(r-1\right) \nonumber \\ 
        &\phantom{=} \times r^{\beta-1} \bigg\{ 
           \, -\frac{3}{\pi^2}\, \left(\left(\frac{1}{\varepsilon}-\log (2\pi) \right) \, \zeta_{4} + \zeta'_{4} \right) \, \left[1+ r^{-1}\right] \bdelta(i(\beta-3)) \nonumber \\
        &\phantom{=} + \frac{1}{12\pi^3} \left( -2\pi^2 \left[ r^{-2}  + r^{-1} \right]\zeta_{3}  +   3 \left[r^{-2} + 4r^{-1} \right]\zeta_5  \right)  \bdelta(i(\beta-4))+\cdots  
        \bigg\} \nonumber \\ 
        &\phantom{=} + \mathcal{O}\left(\varepsilon^{1}\right).
\end{align}
It is noteworthy that the first non-vanishing loop correction is located by the residue at the pole $\beta=3$, shifted by exactly one unit when compared to first term in the stringy corrections at tree level (\ref{eq:StringyCorrectionTermsTreeLevel}).   

\subsection{One Loop Non-Planar Amplitude}
\label{sec:NonPlanarLoopLevelCelestialExpansion}

When compared to the planar case, the Mellin transform for the non-planar sector includes an additional factor of $q^{\alpha' s_{12}/4}$ with $q = e^{-2\pi t}$ as seen in (\ref{eq:NonPlanarPunctureIntegral}). The algebraic manipulations are largely the same, the main difference being that the remaining energy integral now takes the form of a  $\Gamma$-function. Concretely, we have 
\begin{align}
    \mathscr{J}_{\text{NP}}(n,m) &= \frac{g_{10}^2}{\left(\alpha'\right)^{3}} \prod_{i=1}^{4} \int_{0}^{\infty} \dd{\omega_{i}} \omega_{i}^{\Delta_{i}-1} \, \mathcal{A}_{\text{YM}}^{\text{tree}}\left(1^{-},2^{-},3^{+},4^{+}\right) (\alpha' s_{12})^{n+1}(\alpha' s_{23})^{m+1} e^{-\alpha' \pi t \,  s_{12}/2} \nonumber \\ 
    &= 4 (\alpha')^{n+m-1} \frac{(z_{12}\bar{z}_{12})^{n+1}(z_{23}\bar{z}_{23})^{m}}{(z_{14}\bar{z}_{14})} \frac{z_{12}^3}{z_{23}z_{34}z_{41}} \delta(r-\bar{r})\theta(r-1)    \nonumber \\  
     &\phantom{=} \times \prod_{i=1}^{4} \int_{0}^{\infty} \dd{\omega_{{i}}} \prod_{j=1}^{3} \delta\left(\omega_j - \chi_{j} \omega_{4}\right)   \omega_{1}^{\Delta_1 + n + 1} \omega_{2}^{\Delta_2 + n + m + 2} \omega_{3}^{\Delta_3 + m - 1}\omega_{4}^{\Delta_{4}-3} e^{-\alpha' \pi t \,  z_{12}\bar{z}_{12}\omega_{1}\omega_{2}/2} \nonumber \\ 
     &= 4 (\alpha')^{n+m-1}  \frac{(z_{12}\bar{z}_{12})^{n+1}(z_{23}\bar{z}_{23})^{m}}{(z_{14}\bar{z}_{14})} \frac{z_{12}^3}{z_{23}z_{34}z_{41}} \delta(r-\bar{r})\theta(r-1)   \nonumber \\ 
     &\phantom{=} \times  \chi_{1}^{\Delta_1+n+1}\chi_{2}^{\Delta_2+n+m+2}\chi_{3}^{\Delta_{3}+m-1} \int_{0}^{\infty} \dd{\omega_{4}} \omega_{4}^{\Delta+2(n+m)-1}  e^{-\alpha' \pi t \,  z_{12}\bar{z}_{12}\chi_{1}\chi_{2} \, \omega_4^2/2}   \nonumber \\ 
     &= 2 \, g_{10}^2 (\alpha')^{\beta-3} \, K\left(h_i,\bar{h}_i\right) r^{\frac{5-\beta}{3}} \left(r-1\right)^{\frac{2-\beta}{3}} \delta(r-\bar{r})\theta(r-1)  \nonumber \\ 
      &\phantom{=} \times r^{\beta-1-m}   (2\pi t)^{\beta-2-m-n} \, \Gamma\left(2 + m + n - \beta \right),
      \label{eq:NonPlanarEnergyIntegral}
\end{align}
again obtaining the $(\alpha')^{\beta-3}$ factor. 

The remaining computation amounts to performing the moduli space integrals. However, in the non-planar case the $\varepsilon$-regularisation is not needed for the $q$ and thereby $t$-dependent parts of the eMZVs, since the additional factors of the string moduli parameter $t$ act as natural regulator, yielding integrals of the form
 \begin{align}
 	   \int_{0}^{\infty} \dd{t}  (2\pi t)^{\beta-2-m-n} \, \Phi\left(a,b;q\right) &= \sum_{k,\ell \geq 1} \frac{1}{k^a \ell^b} \int_{0}^{\infty} \dd{t}  (2\pi t)^{\beta-2-m-n} \, e^{-2 k \ell \pi \, t} \nonumber \\
 	   &= \frac{1}{2\pi} \Gamma\left(\beta-1-m-n\right)  \zeta_{\beta-1-m-n+a} \zeta_{\beta-1-m-n+b}.
 	   \label{eq:RegIntegralNP}  
 \end{align}
Combining this $\Gamma$-function with the one in the non-planar energy integral (\ref{eq:NonPlanarEnergyIntegral}) yields 
\begin{align}
	\Gamma\left(2 + m + n - \beta \right)\Gamma\left(\beta-1-m-n\right) = \frac{\pi}{\sin \pi \left(m+n-\beta\right)} = \left(-1\right)^{1+m+n}\frac{\pi}{\sin \pi\beta},
\end{align}
which is the same sine factor present in the stringy corrections of the tree level result (\ref{eq:TreeEnergyIntegral2}). The two domains of convergence of this and the previous $\Gamma$-function integral in (\ref{eq:NonPlanarEnergyIntegral}) can be summarised as $ 1 + m + n < \real{\left(\beta\right)} < 2 + m + n $. This implies that both zeta functions in (\ref{eq:RegIntegralNP}) are evaluated in their domain of absolute convergence unless either $a<1$ or $b<1$. However, in these cases and for the purpose of a more general analysis, one can utilize the well known analytic continuation of the zeta function to a meromorphic function on the entire complex plane 
\begin{align}
	\zeta_{s} = 2^{s}\pi^{s-1}\sin\left(\frac{\pi s}{2}\right) \Gamma(1-s)\zeta_{1-s}, \quad \forall \, s\in \mathbb{C} - \{1\},
	\label{eq:AnalyticalContinuationZeta}
\end{align}
except for $s=1$ where it retains a simple pole with residue $1$. When integrating over $q$-independent parts of the eMZVs, the integral will simply yield a generalised delta distribution once again
\begin{align}
	 \int_{0}^{\infty} \dd{t}  (2\pi t)^{\beta-2-m-n} = \frac{1}{2\pi} \int_{0}^{\infty} \dd{u}  u^{\beta-2-m-n} = \bdelta(i(1+n+m-\beta)).
\end{align} 
The $\Gamma$-functions  $\Gamma\left(2 + m + n - \beta \right) $ that were previously obtained via Mellin transform in (\ref{eq:NonPlanarEnergyIntegral}) actually trivialise on the support of $\bdelta(i(1+n+m-\beta))$ to $\Gamma\left(1 \right) = 1 $. 

Starting from the non-planar string amplitude expansion (\ref{eq:NonPlanarPunctureIntegral}) and assembling all the terms, we ultimately obtain 
\begin{align}
	\tilde{\mathcal{A}}_{\text{NP}}\left(1,2,3,4\right) &= 2 \, g_{10}^2 (\alpha')^{\beta-3} \, K\left(h_i,\bar{h}_i\right) r^{\frac{5-\beta}{3}} \left(r-1\right)^{\frac{2-\beta}{3}} \delta(r-\bar{r})\theta(r-1) \nonumber \\ &\phantom{=} \times r^{\beta-1} \bigg\{
		\bdelta\left(i\left(\beta-1\right)\right) + \frac{1}{2} \, \zeta_{\beta+1}\left[
	\bdelta\left(i\left(\beta-3\right)\right) - \bdelta\left(i\left(\beta-4\right)\right)
	\right]
	  \nonumber \\
	&\phantom{=} -    \frac{5}{2}  \frac{\pi}{\sin\left(\pi \beta\right)} \, \zeta_{\beta-1} \left(
	 \,  \zeta_{\beta-4} \left[1+r^{-1}+r^{-2}\right] 
	 - \zeta_{\beta-6} \left[r^{-1}+r^{-2}\right] 
	\right) + \cdots
	\bigg\}.
	\label{eq:CelestialNPExpansionFullyIntegrated}
\end{align}
The second line of (\ref{eq:CelestialNPExpansionFullyIntegrated}) contains the usual expected poles on the positive real axis resulting from the IR expansion of the amplitudes in terms of Mandelstam variables. A noteworthy novelty here is the first pole is located at $\beta = 1$, that is to the left of the first pole of all other loop expansions and even the stringy tree corrections starting at $\beta =2$, yet still to the right of the YM pole at $\beta = 0$. However, the the third line contains the same $\sin\left(\pi \beta\right)^{-1}$ factor that previously encountered in the full tree level result (\ref{eq:TreeEnergyIntegral2}), giving rise to potential simple poles at $\beta = n, \, \forall \, n \in 
\mathbb{Z}$. In particular, this allows for the existence of poles positioned on the negative real line that correspond to UV sector contributions. However, here we can easily show that the residues at the poles actually vanish because of the structure of the non-planar string amplitude expansion. Concretely, the zeta function possesses trivial zeros $\zeta_{n} = 0 \,  $ for all even $n \in \mathbb{Z}_{<0}$ as seen from the $\sin\left(\pi s/2\right)$ factor in its analytic continuation (\ref{eq:AnalyticalContinuationZeta}). This implies that the accompanying products of zeta functions have trivial zeros for all $ \beta=n\in \mathbb{Z}_{\leq 0} $, resulting in vanishing residues on all negative integers including zero
\begin{align}
	\Res\left[ \, \frac{\zeta_{\beta-1}\zeta_{\beta-4}}{\sin\left(\pi \beta\right)} \right] = \Res\left[ \, \frac{\zeta_{\beta-1}\zeta_{\beta-6}}{\sin\left(\pi \beta\right)} \right] = 0, \quad \forall \, \beta=n\in \mathbb{Z}_{\leq 0}.
\end{align}
Let us complete this discussion by also considering the residues of these terms along the positive real axis. Besides the simple poles located at every positve integer value $\beta=n\in \mathbb{Z}_{>0}$ coming from the inverse sine function, there are additional simple poles where the argument of any of the zeta functions is equal to one. All these cases can easily be computed via residue theorem. Overall, we find 
\begin{align}
    \Res\left[ \, \frac{\pi \zeta_{\beta-1}\zeta_{\beta-4}}{\sin\left(\pi \beta\right)}, \beta = n \in \mathbb{Z} \right] = \begin{cases}
			\zeta'_{-2}, & \beta = 2 \\
            -\gamma \, \zeta_4 - \zeta'_{4}, & \beta = 5 \\ 
            (-1)^{n} \zeta_{n-1}\zeta_{n-4}, & \text{else},
		 \end{cases}
\end{align}
where $\gamma$ is the Euler–Mascheroni constant. Analogously, one can compute 
\begin{align}
    \Res\left[ \, \frac{\pi\zeta_{\beta-1}\zeta_{\beta-6}}{\sin\left(\pi \beta\right)}, \beta = n \in \mathbb{Z}\right] = \begin{cases}
			\zeta'_{-4}, & \beta = 2 \\
            -\gamma \, \zeta_6 - \zeta'_{6}, & \beta = 7 \\ 
            (-1)^{n} \zeta_{n-1}\zeta_{n-4}, & \text{else}.
		 \end{cases}
\end{align}
A compact all-order $\alpha'$-expansion at loop level would be required to generalise this analysis.           

\subsection{String Moduli Integrals evaluated at integer values of \texorpdfstring{\boldmath $\beta$}{beta} }
\label{sec:RemainingIntegralEvaluated}

Throughout this section, we have first used known $\alpha'$-expansions of the one loop superstring amplitudes and applied the Mellin transform afterwards. This allowed us to unveil at least part of the pole structure in the complex $\beta$-plane at positive integers. We would like to display a connection between these results and the remaining string moduli integral $\Psi_{\text{cyl}}\left(\beta\right)$ introduced at the end of the previous section in (\ref{eq:RemainingIntegralBeta}). Concretely, we would like to evaluate it at integers values in $\beta$ and compare it with the corresponding residues in the planar cylinder case displayed in (\ref{eq:ExpansionPlanarCylinder}), expecting matching results. 

The first pole of the cylinder contribution is positioned at $\beta=2$. This is precisely the value where $\Psi_{\text{cyl}}\left(\beta\right)$ becomes trivial, as the integrand simplifies to $1$. Thus, we simply compute    
\begin{align}
 	\Psi_{\text{cyl}}\left(2\right) =  \int_{0}^{1} \frac{\dd{q}}{q} \, \int_{1234} \, =   \int_{0}^{1} \frac{\dd{q}}{q} \,  \frac{1}{6} = \int_{0}^{1} \frac{\dd{q}}{q} \, \varphi(0,0,0),  
\end{align}
obtaining exactly the same residue at $\beta=2$ as in the expansion (\ref{eq:ExpansionPlanarCylinder}).\footnote{Note the extra factor of $2\pi$ coming from the definition of the delta distribution.} 

For $\beta=3$ the integrand is just raised to unit power, making it comparatively easy to compute by brute force, at least when consulting the integral software of your choice. It involves evaluating the iterative integral of the string worldsheet puncture positions $\int_{1234}$ over several copies of $4 \sin^2{(\pi m x_{ij})} \frac{q^{mn}}{m}$ for $(i,j)\in\{1,2,3,4\}$ where $i<j $ and $(n,m)\in \mathbb{N}_{\geq 1}$, as well as the the $\log[\sin(\pi x_{ij})]$ terms. Ultimately, one obtains
\begin{align}
	\Psi_{\text{cyl}}\left(3\right)= \int_{0}^{1} \frac{\dd{q}}{q} \, \left[1+r^{-1}\right] \left( \frac{3}{\pi^2} \sum_{m,n=1}^{\infty} \frac{1}{m^3} q^{mn} +  \frac{\zeta_{3}}{4\zeta_{2}} \right)	 = \int_{0}^{1} \frac{\dd{q}}{q} \, \left[1+r^{-1}\right] 2 \, \varphi\left(0,1,0,0\right),
	\label{eq:ComputerResultAtBeta3}  
\end{align} 
reproducing the same term as in (\ref{eq:ExpansionPlanarCylinder}) yet again. However, this computational approach is just as unsatisfyingly opaque as it is limited, as even the next pole $\beta=4$ turns out to be too challenging, at least for Mathematica.\footnote{The mixed $\sin\cdot \log$ as well as the squared logarithmic terms $\log^2 $ yielded no closed form result.} It is therefore of utmost importance, for both practical and conceptual reasons, to develop a more systematic framework to evaluate these integrals. 

Fortunately, this is precisely what has been done already in \cite{broedelEllipticMultipleZeta2015}. Indeed, this is how the expansion of the string formfactor in terms of eMZVs (\ref{eq:PlanarPunctureIntegral}), which we subsequently recast in the celestial basis, was derived in the first place. The main difference here is that, rather than using the series representation of the exponential in the integrand of (\ref{eq:BasicIterativeIntegral})
\begin{align}
	I_{1234}(q) =  \int_{1234} \, \prod_{i<j}^{4} \, \sum_{n=1}^{\infty} \frac{1}{n!}\left(\alpha' s_{ij}P_{ij}\right)^n,
\end{align}
and subsequently gathering all the powers of the Green functions $P_{ij}$ before performing the $\int_{1234}$ integral, we already performed the Mellin transform, leading to a multinomial that for $\beta \in \mathbb{N}_{\geq2}$ can be expressed as
\begin{multline}
		\Psi_{\text{cyl}}\left(\beta\right) =  \int_{0}^{1} \frac{\dd{q}}{q} \, \int_{1234} \, \sum_{k_{1},\ldots,k_{8}\geq 0} \binom{\beta-2}{k_{1},\ldots,k_{8}} \left(-1\right)^{(k_3 + k_4 + k_7 +k_8)}   \\
		\times r^{-\left(k_5+k_6 + k_{7} +k_{8}\right)} P_{13}^{k_{1}+k_{5}}P_{24}^{k_{2}+k_{6}} P_{12}^{k_3}P_{34}^{k_4}P_{14}^{k_7}P_{23}^{k_{8}}.
\end{multline}
Nevertheless, this expression can still be handled by the same techniques developed in the original paper,  which we will follow closely in the subsequent analysis.

A central observation is that the iterated integral over $N$ string punctures  (\ref{eq:IntegralOverPunctures})  is invariant under cyclic shifts $z_{i} \to z_{i+1 \!\!\!\mod{N}} $ and reflections $z_{i} \to z_{N-1-i}$ up to a sign $(-1)^N$. This allows us to relate the integrals over the Green functions $P_{ij}$ to one another, drastically reducing the amount of integrals one needs to compute. 

As an introductory example, let us redo the case $\beta=3$. We have a total of $6$ possible $\int_{1234}$ integrals over Green functions $P_{ij}$, $(i,j)\in \mathbb{N}_{\geq 1}$ with $i<j$ that will make an appearance. Using cyclic and reflection symmetry, one can reduce this set to only $2$ basic integrals 
\begin{align}
	\int_{1234} \, P_{12} = \int_{1234} \, P_{23} = \int_{1234} \, P_{34} = \int_{1234} \, P_{14},  \qquad  \int_{1234} \, P_{13} = \int_{1234} \, P_{24},
	\label{eq:IntegralRelationsLinear}
\end{align}   
that can explicitly be expressed in terms of eMZVs
\begin{align}
	\int_{1234} \, P_{12} = \varphi(1,0,0,0), \qquad \int_{1234} \, P_{13} = \varphi(1,0,0,0) + \varphi(0,1,0,0).
	\label{eq:LinearMZVs}
\end{align}
Reintroducing the explicit combination of Green functions by plugging (\ref{eq:GreenFunctionCombinationPlanar}) into (\ref{eq:RemainingIntegralBeta}), simplifying the total integrand using (\ref{eq:IntegralRelationsLinear}) and (\ref{eq:LinearMZVs}), we have 
\begin{align}
	\Psi_{\text{cyl}}\left(3\right) = 2\int_{0}^{1} \frac{\dd{q}}{q} \, \left[1+r^{-1}\right] \int_{1234} \, \left(
		P_{13}-P_{12}
		\right)
	= 2\int_{0}^{1} \frac{\dd{q}}{q} \, \left[1+r^{-1}\right] \varphi(0,1,0,0).
\end{align}
 in agreement with (\ref{eq:ComputerResultAtBeta3}). 
 
 Let us now tackle the case for $\beta=4$ with the same approach. The integrand of $\Psi_{\text{cyl}}\left(4\right)$ is $(\mathcal{P}_{1234} + \mathcal{P}_{1432}/r)^2$. Expressed in terms of Green functions $P_{ij}$, one obtains $k=6$ unique Green functions raised to an overall power of $n=2$, resulting in a total number of $\binom{n+k-1}{k-1} = 21 $ terms.\footnote{Two of the four Green functions in each $\mathcal{P}_{1234}$ and $\mathcal{P}_{1432}$ are the same, see (\ref{eq:GreenFunctionCombinationPlanar}).} Exploiting  cyclic and reflection symmetry of the four-point iterative integral once again, these can be reduced to a basis of only 6 independent integrals as seen in Table \ref{tab:EquivalentGreenFunctions}.
 \begin{table}
    \centering
    $\begin{array}{|l|l|}
    \hline
    \text{Base Integrands} & \text{Equivalent Integrands} \\ 
    \hline
    \hline
	P_{12}^2 &  P_{14}^2 , \,  P_{23}^2, \, P_{34}^2  \\ 
	\hline
	P_{13}^2 & P_{24}^2 \\ 
	\hline
	\hline
	P_{12}P_{13} & P_{12}P_{24}, \, P_{13}P_{14}, \, P_{13}P_{23}, \, P_{13}P_{34}, \, P_{14}P_{24}, \, P_{23}P_{24}, \, P_{24}P_{34} \\ 
	\hline
	P_{12}P_{14} & P_{12}P_{14}, \, P_{14}P_{34}, \,P_{23}P_{34}   \\
	\hline
	P_{12}P_{34} & P_{14}P_{23} \\ 
	\hline  
	P_{13}P_{24} & \\ 
	\hline 
    \end{array}$
    \caption{Integrands that yield equal results within $\int_{1234}$ via cyclic shifts \& reflections }
    \label{tab:EquivalentGreenFunctions}
\end{table}
  Moreover, one can actually use the same symmetries for the five-point iterative integral $\int_{12345}$ to derive further relations for the four-point case by inserting a derivative operator between two Green functions and integrating by parts.\footnote{See equation (4.15) in \cite{broedelEllipticMultipleZeta2015} for an example.} The additional constraints read 
\begin{align}
	\int_{1234} \, P_{12}P_{34} = \int_{1234} \, P_{12}P_{14}, \qquad 2 \int_{1234} \, P_{12}P_{13} =  \int_{1234} \, P_{12}P_{14} +  \int_{1234} \, P_{13}P_{24}, 
\end{align}
further reducing the number of basic integrals to $4$. Before proceeding, we want to note some relations between different eMZVs that can be derived by combining the reflection properties
\begin{align}
	\varphi\left(n_{1},n_{2},\ldots,n_{k-1},n_{k}\right) = \left(-1\right)^{\sum_{i=1}^{k} n_{i}}\varphi\left(n_{k},n_{k-1},\ldots,n_{2},n_{1}\right),
	\label{eq:ReflectionPropertyMZV}
\end{align}
with the shuffle relation
\begin{align}
\varphi\left(n_{1},\ldots,n_{k}\right)\varphi\left(m_{1},\ldots,m_{\ell}\right) = \varphi\left(\left(n_{1},\ldots,n_{k}\right)\shuffle \left(m_{1},\ldots,m_{\ell}\right)\right),
	\label{eq:ShuffleeMZVs}  
\end{align}
at the level of the eMZVs. As is often the case, some of the most useful identities are elaborate ways of expressing zero, and the following ones are no exception. By the reflection property (\ref{eq:ReflectionPropertyMZV}) it immediately follows that $\varphi(1) = 0$. Combining this with evaluations of terms like $\varphi(1)\varphi\left(1,0,0,0\right)$ using the shuffle relation (\ref{eq:ShuffleeMZVs}), one obtains a set of relations
\begin{align}
	2 \varphi\left(1,1,0,0,0\right) + \varphi\left(1,0,1,0,0\right) + \varphi\left(1,0,0,1,0\right) + \varphi\left(1,0,0,0,1\right) &= 0, \label{eq:eMZVShuffleAndReflectionSum1} \\ 
	\varphi\left(1,0,1,0,0\right) + 2\varphi\left(0,1,1,0,0\right) + \varphi\left(0,1,0,1,0\right) + \varphi\left(0,1,0,0,1\right) &= 0, \\ 
	\varphi\left(1,0,0,1,0\right) + \varphi\left(0,1,0,1,0\right) + 2\varphi\left(0,0,1,1,0\right) + \varphi\left(0,0,1,0,1\right)&= 0, \\ 
	\varphi\left(1,0,0,0,1\right) + \varphi\left(0,1,0,0,1\right) + \varphi\left(0,0,1,0,1\right) + 2\varphi\left(0,0,0,1,1\right)&= 0. 
	\label{eq:eMZVShuffleAndReflectionSum4}
\end{align}
Expanding the multinomial in the integrand, expressing all integrals over the string puncture positions in terms of the $4$ basic ones, performing algebraic manipulations using the eMZV relations (\ref{eq:eMZVShuffleAndReflectionSum1})-(\ref{eq:eMZVShuffleAndReflectionSum4}), we obtain 
\begin{align}
	\Psi_{\text{cyl}}\left(4\right) &=   \int_{0}^{1} \frac{\dd{q}}{q} \,  \int_{1234} \, \bigg\{ 2\left[1+r^{-2}\right]P_{12}^2   + 4 r^{-1}P_{12}P_{14} + 2\left[1+2r^{-1}+r^{-2}\right]\left(P_{13}^2 -2P_{12}P_{13}\right)  \bigg\}  \nonumber \\ 
	&= 4  \int_{0}^{1} \frac{\dd{q}}{q} \,   \bigg\{
	\left[1+r^{-2}\right] \varphi\left(1,1,0,0,0\right) - r^{-1}\varphi\left(1,0,0,0,1\right)  \nonumber \\ 
	&\phantom{=4  \int_{0}^{1} \frac{\dd{q}}{q} \,   \bigg\{}   + \left[1+2r^{-1}+r^{-2}\right]\left(\varphi\left(0,1,1,0,0\right)-\varphi\left(1,1,0,0,0\right)\right) 
	\bigg\} \nonumber \\ 
	&= 4  \int_{0}^{1} \frac{\dd{q}}{q} \, \bigg\{ 
	\left[1+r^{-2}\right] \varphi\left(0,1,1,0,0\right) - r^{-1}\varphi\left(0,1,0,1,0\right) \nonumber \\ 
	&\phantom{=4  \int_{0}^{1} \frac{\dd{q}}{q} \,   \bigg\{} 
	+ r^{-1}\left(\varphi\left(1,0,0,1,0\right) - \varphi\left(0,1,0,0,1\right)\right)
	\bigg\} \nonumber \\ 
	&= 4  \int_{0}^{1} \frac{\dd{q}}{q} \, \bigg\{ 
	\left[1+r^{-2}\right] \varphi\left(0,1,1,0,0\right) - r^{-1}\varphi\left(0,1,0,1,0\right)\bigg\},
\end{align}
where we used $\varphi\left(1,0,0,1,0\right) =  \varphi\left(0,1,0,0,1\right)$ as implied by the reflection property (\ref{eq:ReflectionPropertyMZV}) to cancel the second line after the third equals sign. This is again consistent with the residue of the $\beta=4$ pole in (\ref{eq:ExpansionPlanarCylinder}).

\section{Conclusion and Outlook}
\label{sec:Conclusions}

In this paper, we transformed all sectors of the type \MakeUppercase{\romannumeral 1} open superstring amplitude at one-loop to conformal correlators on the celestial sphere, dubbed celestial string integrands, emphasising the remaining integrals over the string worldsheet moduli space. We find that all sectors yield similar celestial integrands, with the planar (non-)orientable sector differing solely by an overall constant $-32/N_{G}$ and a sign shift $q \to -q$ in the exponentiated moduli parameter $q = e^{-2\pi t}$ and the non-planar orientable sector allowing for punctures on both boundaries of the string worldsheet. This property is inherited by regular superstring scattering amplitudes, due to the fact that the Mellin transform with respect to the energies of the scattered strings does not interact with the string worldsheet moduli. 

We presented an explicit formula (\ref{eq:RemainingIntegralBeta}) for $\Psi_{\text{cyl}}\left(\beta\right)$, the remaining moduli space integral over the celestial integrand in the case of the planar cylinder contribution. We proposed methods to expand the celestial integrand into a form of products of genus one Green functions, each raised to some individual, possibly complex and/or negative power. This represents a generalisation of the expansion performed in \cite{broedelEllipticMultipleZeta2015}, where only non-negative integer powers occurred. As such, a solution for a general complex parameter $\beta \in \mathbb{C}$   will most likely require a similar systematic analysis in terms of special functions such as eMZVs or a generalization thereof. We leave this to future work. 

In order to gain further insights into the celestial superstring amplitudes regardless, we employed their $\alpha'$-expansions   \cite{Green:1981xx,broedelEllipticMultipleZeta2015,broedelTwistedEllipticMultiple2018b}. Recasting these expressions into the celestial basis unveiled the pole structure of the amplitudes in the complex $\beta$-plane, which is related to the conformal scaling dimension of the CCFT operators corresponding to the scattered string states. The analysis was performed both at tree level, allowing for a proof of concept via direct comparison with the known closed-form expression \cite{stiebergerStringsCelestialSphere2018}, and then extended to all sectors of the open superstring amplitude at one-loop level. The poles resulting from the $\alpha'$-expansions were found to exclusive reside at positive integer values.

At tree level, the residues associated to each pole were shown to consist of powers of the conformal cross ratio $r$ accompanied by zeta values $\zeta_{n}, n\in\mathbb{N}_{\geq 2}$, which arise as evaluations of the more general MZV structure $\zeta\left(\beta-k,\{1\}^{k}\right)$ of the full result at the positions of the poles $\beta = n\in\mathbb{N}_{\geq 2}$. The additional, potentially divergent, $\int_{0}^{1} \dd{q}/q$ integral over the exponentiated moduli parameter $q$ present at loop-level were evaluated for all sectors with the provided $\varepsilon$-regularization method. In doing so, the full residue structure consisting of eMZVs at the level of the $\int_{0}^{1} \dd{q} $-integrand, reduced to that of zeta values in a similar fashion to the tree level case. The only exception to this is a part of the non-planar sector. While all other residues can easily be read off as the coefficients of the accompanying generalized delta distributions \cite{donnayAsymptoticSymmetriesCelestial2020,panoDistributionalCelestialAmplitudes2024} marking the position of the poles, part of the non-planar sector consisted of meromorphic functions, requiring additional analysis of the pole structure. This was achieved using the analytic continuation of the zeta function  to the entire $\beta$-plane and applying the residue theorem, again reducing the coefficients to simple zeta values or derivatives thereof. Furthermore, these terms exhibited an interesting phenomenon, where the products of zeta functions can be shown to possess trivial zeros at all non-positive integers, resulting in the annihilation of potential poles at these positions indicated by the inverse sine function $\sin{\left(\pi\beta\right)}^{-1}$ present in all terms. Those are generally indicative of UV contributions and their disappearance is expected from the soft UV behaviour of quantum gravity theories \cite{arkani-hamedCelestialAmplitudesUV2021}. However, in this case this seems to be connected to the nature of a low-energy $\alpha'$-expansion in that it is unable to probe the UV. In the same spirit, an analytic continuation of the fully fledged tree level celestial string amplitude was provided. This revealed an additional infinite set of poles on the real line located at every integer in the region $\left(-\infty, k+1 \right] \in \real{\left(\beta\right)} $ corresponding to the $k$-th stringy correction term to the YM amplitude. However, in this case the residues in the left half-plane $\real{\left(\beta\right)} < 0 $ do not seem to vanish, at least identically, as the MZVs do not possess any trivial zeros there \cite{zhaoAnalyticContinuationMultiple2000}. This is indeed what one would expect from a fully fledged celestial amplitude which, by their very definition as an integral over all energies of a flat-space amplitude, are sensitive to the UV. 

In all cases and with both approaches, we found that the $\alpha'$-dependency of the celestial string amplitudes simplifies to an overall factor $\left(\alpha'\right)^{\beta}$ at tree, and   $\left(\alpha'\right)^{\beta-3}$ at loop level, consistent with previous results \cite{stiebergerStringsCelestialSphere2018,donnayCelestialOpenStrings2023a}. 

Finally, we evaluated the integral over the celestial cylinder amplitude integrand  $\Psi_{\text{cyl}}\left(\beta\right)$ for $\beta = 2,3,4$ and demonstrated that it yields the same residues as obtained with the $\alpha'$-expansion ansatz. Besides providing a consistency check, this demonstrates that the iterative integral over the string worldsheet punctures $\int_{1234}$ commutes with the Mellin transform. 

There is still a lot of work to be done. In the following, we want to briefly discuss some of the remaining questions. This list is not meant to be exhaustive by any means. 

The arguably most important challenge connected to this paper is the derivation of a fully-integrated, non-expanded version of the celestial superstring amplitude at one-loop. Solving the integral $\Psi_{\text{cyl}}\left(\beta\right)$, alongside its non-orientable and non-planar cousins, for generic $\beta \in \mathbb{C}$ represents one possible pathway for achieving this goal. 

In this work we exclusively dealt with uncompactified, ten-dimensional spacetime.  Another potentially interesting research avenue is to consider different compactifaction scenarios for the celestial superstring and investigate how these will affect the celestial amplitude. A simple supersymmetry preserving choice is the torus. As pointed out in \cite{broedelEllipticMultipleZeta2015}, wrapping up any of the extra dimensions on circular dimensions of radius $R$  would lead to additional  $ \sum_{n= -\infty}^{\infty}e^{-\pi t (nR)^2/\alpha'}$ terms in the overall integrand for each dimension compactified this way. Note that these terms will only affect the moduli parameter integral $\int_{0}^{\infty} \dd{t}$  while the integral over the string worldsheet positions remains unaffected. 

Last but not least, this work dealt solely with the analytic part of the string amplitude. Generally, there are non-analytic contributions arising from the so-called threshold singularities that originate from the exchange of massless states in the string loop. These manifest as discontinuities in the Mandelstam variables $s_{ij}$, represented by $\log(s_{ij})$ terms. It would be interesting to see how these terms can be recast in the celestial basis.  We plan to address this in a future work.

\acknowledgments

Daniel Bockisch would like to thank Oliver Schlotterer, Ivo Sachs and Stephan Stieberger for helpful correspondence throughout the creation of this paper. DB is supported by the Excellence Cluster Origins of the DFG under Germany's Excellence Strategy EXC-2094 390783311. 

\appendix

\section{Explicit eMZVs}
\label{sec:eMZVs}

The elliptical multiple zeta values can generally be expressed as combination of zeta values and infinite series over the exponentiated torus moduli parameter $q=e^{-2\pi t}$. In the following we list the explicit form of all eMZVs appearing in the final results of our computations 

\begin{align}
     \varphi(0,0,0) &= \frac{1}{6}, \\
     \varphi\left(0,0,2\right) &= -\frac{\zeta_2}{3} + 2\sum_{m,n\geq 1}^{\infty} \frac{m}{n^2} q^{mn},  \\ 
     \varphi(0,1,0,0) &= \frac{\zeta_{3}}{8\zeta_{2}} + \frac{3}{2\pi^2}\sum_{m,n\geq 1}^{\infty}  \frac{1}{m^3} q^{mn},
    \\
    \varphi\left(0,1,0,1\right) &= \frac{3\zeta_3}{4\pi^2} + \frac{3}{2\pi^2} \sum_{m,n\geq 1}^{\infty} \frac{1}{n^3}q^{mn}, \\
    \varphi\left(0,3,0,0\right) &= -3 \sum_{m,n\geq 1}^{\infty} \frac{m^2}{n^3}q^{mn}, \\  
    \varphi(0,1,1,0,0) &= \frac{\zeta_{2}}{15} - \frac{1}{2\pi^2}\sum_{m,n\geq 1}^{\infty} \frac{n}{m^4} q^{mn} + \frac{1}{3}\sum_{m,n\geq 1}^{\infty} \frac{n}{m^2} q^{mn},
    \\ 
     \varphi(0,1,0,1,0) &= -\frac{\zeta_{2}}{60} + \frac{2}{\pi^2}\sum_{m,n\geq 1}^{\infty} \frac{n}{m^4} q^{mn} - \frac{1}{3}\sum_{m,n\geq 1}^{\infty} \frac{n}{m^2} q^{mn}. 
\end{align}

\bibliography{references} 
\bibliographystyle{JHEP}

\end{document}